\documentclass[11pt,a4paper]{article}
\usepackage{jheppub}
\usepackage{atlasphysics}
\usepackage{multirow}
\usepackage{euscript}

\usepackage{preprintcover} 
\PreprintCoverPaperTitle{Search for dark matter candidates and large
  extra dimensions in events with a jet and missing transverse
  momentum with the ATLAS detector} 
\PreprintIdNumber{CERN-PH-EP-2012-210} 
\PreprintCoverAbstract{A search for new phenomena in events with a
  high-energy jet and large missing transverse momentum is performed
  using data from proton-proton collisions at $\sqrt{s}=7\TeV$ with
  the ATLAS experiment at the Large Hadron Collider. Four kinematic
  regions are explored using a dataset corresponding to an integrated
  luminosity of $4.7\ifb$. No excess of events beyond expectations
  from Standard Model processes is observed, and limits are set on
  large extra dimensions and the pair production of dark matter
  particles.}  
\PreprintJournalName{JHEP} 

\def\ptmiss{\ensuremath{\mathbf{p}_{\mathrm{T}}^{\mathrm{miss}}}}
\def\jettwo{\ensuremath{\mathbf{p}_{\mathrm{T}}^{\mathrm{jet2}}}}
\def\jetthree{\ensuremath{\mathbf{p}_{\mathrm{T}}^{\mathrm{jet3}}}}

\title{Search for dark matter candidates and large extra dimensions in
  events with a jet and missing transverse momentum with the ATLAS
  detector}

\collaborationImg{\includegraphics{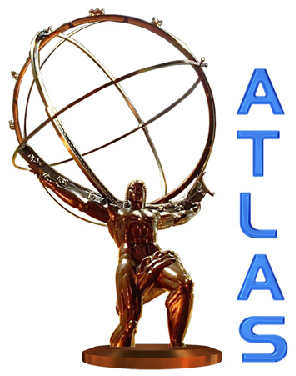}}

\collaboration{The ATLAS Collaboration}

\emailAdd{Atlas.Publications@cern.ch}

\abstract{A search for new phenomena in events with a high-energy jet
  and large missing transverse momentum is performed using data from
  proton-proton collisions at $\sqrt{s}=7\TeV$ with the ATLAS
  experiment at the Large Hadron Collider. Four kinematic regions are
  explored using a dataset corresponding to an integrated luminosity
  of $4.7\ifb$. No excess of events beyond expectations from Standard
  Model processes is observed, and limits are set on large extra
  dimensions and the pair production of dark matter particles.}

\keywords{Hadron-Hadron Scattering}

\begin{document}
\maketitle
\flushbottom
\clearpage

\section{Introduction}
\label{sec:intro}
Event topologies with a single jet with large transverse energy and
large missing transverse momentum, referred to as \emph{monojets} in
the following, are important final states for searches for new
phenomena beyond the Standard Model (SM) at a hadron collider. The
primary SM process that results in a true monojet final state is
$Z$-boson production in association with a jet, where the $Z$ boson
decays to two neutrinos. A further important reducible contribution to
this final state consists of events that include a $W$ boson and a
jet, where the charged lepton from the $W$-boson decay is not
reconstructed.

Phenomenological scenarios beyond the Standard Model (BSM) that result
in a monojet final state include
supersymmetry~\cite{Miyazawa:1966,Ramond:1971gb,Golfand:1971iw,Neveu:1971rx,Neveu:1971iv,Gervais:1971ji,Volkov:1973ix,Wess:1973kz,Wess:1974tw,Carena:2008mj,Allanach:2010pp}
and large extra dimensions (LED)~\cite{ArkaniHamed:1998rs}. A
model-independent treatment of the production of dark matter (DM)
particles at the Large Hadron Collider (LHC) has been proposed
recently, where DM particles are pair-produced in association with a
jet~\cite{Beltran:2010ww,Rajaraman:2011wf,Fox:2011pm}. In the
following, a search for an excess of monojet events over SM
expectations is performed. The results are interpreted in a framework
of LED and DM particle pair production. They are based on a dataset of
$4.7\ifb$ of proton-proton (\emph{pp}) collisions at $\sqrt{s} =
7\TeV$ recorded with ATLAS at the LHC and supersede those presented in the 2010 ATLAS
monojet analysis that used $35\ipb$ of data~\cite{Aad:2011xw}. Other
monojet searches were performed in Run I and Run II at the
Tevatron~\cite{Abazov:2003gp,Abulencia:2006kk,2012PhRvL108u1804A} and
also by CMS with the 2010~\cite{Chatrchyan:2011nd} and
2011~\cite{CmsPreprint} LHC datasets. None of these found evidence of
new phenomena beyond the Standard Model.

Models of large extra spatial dimensions have been proposed to remove
the hierarchy
problem~\cite{Weinberg:1975gm,Gildener:1976ai,Weinberg:1979bn,Susskind:1978ms}
by addressing the weakness of gravity relative to all other
forces. One popular model of LED that is often used to interpret the
results of monojet searches at particle colliders is that of
Arkani-Hamed, Dimopoulos, Dvali (ADD) ~\cite{ArkaniHamed:1998rs}. In
this model, gravity propagates in the $(4 + n)$-dimensional bulk of
space-time, while the SM fields are confined to four dimensions. The
large observed difference between the characteristic mass scale of
gravity (the Planck mass) and the electroweak scale (as characterised
by the $W$-boson mass) is the result of the four-dimensional
interpretation of the Planck scale,
$M_{\mathrm{Pl}}=1.2\times10^{19}\GeV$, which is related to the
fundamental $(4 + n)$-dimensional Planck scale ($M_\mathrm{D}$) by
${M_{Pl}}^2=8\pi\ {M_\mathrm{D}}^{2+n}R^n$, where $n$ and $R$ are the
number and size of the extra dimensions, respectively. An appropriate
choice of $R$ for a given $n$ results in a value of $M_\mathrm{D}$
close to the electroweak scale.  The extra spatial dimensions are
compactified, resulting in a Kaluza-Klein tower of massive graviton
modes. At hadron colliders, these graviton modes can be produced in
association with a jet. The production processes include $qg
\rightarrow qG, gg \rightarrow gG,$ and $q\bar{q} \rightarrow gG$,
where $G$ stands for the tower of gravitons, $q$ for a quark, and $g$
for a gluon. As gravitons do not interact with the detector, these
processes give rise to a monojet signature~\cite{Giudice:1998ck}.

\emph{Particle dark matter} is a well-established paradigm to explain
a range of astrophysical measurements~(see for example
ref.~\cite{Bertone:2004pz} for a recent review).  Since none of the
known SM particles are adequate DM candidates, the existence of a new
particle is hypothesised, with properties suitable to explain the
astrophysical measurements. One class of particle candidates of
interest for searches at the LHC consists of weakly interacting massive
particles (WIMPs)~\cite{Steigman:1984ac}. These are expected to couple
to SM particles through a generic weak interaction, which could be the
known weak interaction of the SM or a new type of interaction. Such a
new particle is a cold dark matter candidate, which can be produced at the
LHC. It results in the correct relic density values for
non-relativistic matter in the early universe~\cite{Kolb:1990vq}, as
measured by the WMAP satellite~\cite{Komatsu:2010fb}, if its mass lies
in the range between a few GeV and a TeV and if it has
electroweak-scale interaction cross sections. The fact that a new
particle with such properties can be a thermal relic of the early
universe in accordance with the WMAP measurements is often referred to
as the WIMP miracle. Many new particle physics models designed to solve
the hierarchy problem also predict WIMPs.

Because WIMPs do not interact with the detector material, their
production leads to signatures with missing transverse momentum
(\ptmiss)\footnote{Letters in bold font are used for vector
  quantities.}, the magnitude of which is called \met. Searches
involving \met\ at the LHC are therefore canonical WIMP searches,
although the LHC experiments cannot establish whether a WIMP candidate
is stable on cosmological time scales and hence a DM candidate. In
some supersymmetric models, WIMPs are expected to be dominantly
produced in cascade decays of heavier unstable supersymmetric
particles along with high transverse momentum
($\pt=|\mathbf{p}_{\mathrm{T}}|$) SM particles. In a more
model-independent approach, WIMP pair production at colliders is
proposed to yield detectable \met\ if the WIMP pair is tagged by a jet
or photon from initial- or final-state
radiation~(ISR/FSR)~\cite{Birkedal:2004xn,Beltran:2010ww}. Even though
this approach does not rely on a specific BSM scenario, it does have
assumptions: WIMPs are pair-produced at the LHC and all new particles
mediating the interaction between WIMPs and the SM are too heavy to be
produced directly; they can thus be integrated out in an effective
field theory approach. The resulting interaction is hence a contact
interaction between the dark sector and the SM. It is worth noting
that the DM particles are not explicitly assumed to interact via the
weak force. They may also couple to the SM via a new force. Throughout
this work, the terms WIMP and DM particle (candidate) are synonymous.

\begin{table}[htb]
\centering
\begin{tabular}{|c|ccc|}
\hline
Name & Initial state & Type & Operator\\[0.1cm]\hline
D1 &$qq$&scalar&$\frac{m_q}{M^3_\star}\bar{\chi}\chi\bar{q}q$\\[0.4cm]
D5 &$qq$&vector&$\frac{1}{M^2_\star}\bar{\chi}\gamma^{\mu}\chi\bar{q}\gamma_{\mu} q$\\[0.4cm]
D8 &$qq$&axial-vector&$\frac{1}{M^2_\star}\bar{\chi}\gamma^{\mu}\gamma^5\chi\bar{q}\gamma_{\mu}\gamma^5 q$\\[0.4cm]
D9 &$qq$&tensor&$\frac{1}{M_\star^2}\bar{\chi}\sigma^{\mu\nu}\chi\bar{q}\sigma_{\mu\nu}q$\\[0.4cm]
D11&$gg$&scalar& $\frac{1}{4M_\star^3}\bar{\chi}\chi\alpha_s(G^a_{\mu\nu})^2$\\
\hline
\end{tabular}
\caption{Effective interactions coupling Dirac fermion WIMPs to
  Standard Model quarks or gluons, following the formalism of
  ref.~\cite{Goodman:2010ku}. The tensor operator D9 describes a 
  magnetic-moment coupling. The factor of the strong coupling
  constant $\alpha_s$ in the definition of D11 accounts for this
  operator being induced at one-loop level. $G_{\mu\nu}$ is the
  colour field-strength tensor.} 
\label{table:wimp:operators}
\end{table}
It is assumed here that the DM particle is a Dirac fermion $\chi$,
where the only difference for Majorana fermions would be that certain
interaction types are not allowed and that the cross section for each
operator is larger by a factor of four. Five interactions are
considered~(table~\ref{table:wimp:operators}), namely D1, D5, D8, D9,
D11, following the naming scheme of ref.~\cite{Goodman:2010ku}. D1,
D5, D8, and D9 describe different bilinear quark couplings to WIMPs,
$qq\ra \chi\chi$, and D11 describes the process $gg\ra \chi\chi$. The
14 operators for Dirac fermions in ref.~\cite{Goodman:2010ku} fall
into four categories with characteristic \met\ spectral shapes. D1,
D5, D9, and D11 are a representative set of operators for these four
categories, while D8 falls into the same category as D5 but is listed
explicitly in table~\ref{table:wimp:operators} because it is often
used to convert LHC limits into limits on DM pair production. In the
operator definitions in table~\ref{table:wimp:operators}, $M_*$ is the
suppression scale of the heavy mediator particles that are integrated
out. The use of a contact interaction to produce WIMP pairs via heavy
mediators is considered conservative because it rarely overestimates
cross sections when applied to a specific BSM scenario.  Cases where
this approach is indeed optimistic are studied in
refs.~\cite{Fox:2011pm,Friedland:2011za}. The effective theory
provides a useful framework for comparing LHC results to direct or
indirect dark matter searches. Within this framework, interactions of
SM and DM particles are described by only two parameters, the
suppression scale $M_*$ and the DM particle mass $m_\chi$.

\section{Data and simulated samples}
\label{sec:data}
The ATLAS detector~\cite{Aad:2008zzm,Aad:2009wy} at the LHC covers the
pseudorapidity\footnote{ATLAS uses a right-handed coordinate system
  with its origin at the nominal interaction point (IP) in the centre
  of the detector and the $z$-axis along the beam pipe. The $x$-axis
  points from the IP to the centre of the LHC ring, and the $y$-axis
  points upward. Polar coordinates $(r,\phi)$ are used in the
  transverse ($x$,$y$)-plane, $\phi$ being the azimuthal angle around
  the beam pipe.  The pseudorapidity is defined in terms of the polar
  angle $\theta$ as $\eta=-\ln\tan(\theta/2)$.} range of $|\eta| <
4.9$ and all of $\phi$. It consists of an inner tracking detector
surrounded by a thin superconducting solenoid, electromagnetic and
hadronic calorimeters, and an external muon spectrometer incorporating
large superconducting toroidal magnets.  A three-level trigger system
is used to select interesting events for recording and subsequent
offline analysis.  Only data for which all subsystems described above
were operational are used. Applying these requirements to $pp$
collision data, taken at a centre-of-mass energy of $\sqrt{s}=7\TeV$
with stable beam conditions during the 2011 LHC run, results in a data
sample with a time-integrated luminosity of $4.7\ifb$, determined with
an uncertainty of 3.9\%~\cite{Aad:2011dr,confnote:lumi}.

\begin{table}[tbp]
  \begin{center}
    \begin{tabular}{|l|llll|}\hline
      Process & Generator & Parton shower & Underlying event & PDF
      \\\hline
      $Z/W$+jets & {\tt ALPGEN} & {\tt HERWIG} & {\tt JIMMY} & {\tt
        CTEQ6L1}\\
      $t\bar{t}$, single $t$ & {\tt MC@NLO} & {\tt HERWIG} & {\tt
        JIMMY} & {\tt CTEQ6.6}\\
      Di-boson & {\tt SHERPA} & {\tt SHERPA} & {\tt SHERPA} & {\tt CTEQ6L1}\\
      ADD & {\tt PYTHIA} & {\tt PYTHIA} & {\tt PYTHIA} & {\tt
        MSTW2008\,LO$^{**}$} / {\tt CTEQ6.6} \\ 
      WIMPs & {\tt MADGRAPH5} & {\tt PYTHIA} & {\tt PYTHIA} & {\tt CTEQ6L1}\\\hline
    \end{tabular}
    \caption{\label{tab:MC} Overview of the main simulated samples.}
  \end{center}
\end{table}
Monte Carlo (MC) simulations are used both as part of the background
estimation and to model signal processes. Processes that dominate the
background are $Z$- or $W$-boson production in association with jets,
which are simulated with {\tt ALPGEN}~\cite{Mangano:2002ea} using the
parton distribution function (PDF) set {\tt
  CTEQ6L1}~\cite{Pumplin:2002vw}. The $W\ra \mathrm{\ell} \nu$ plus
jets and $Z\ra \nu\bar{\nu}$ plus jets samples are simulated with up
to six additional partons at leading order, while the process
$Z/\gamma^*\ra \ell^+ \ell^-$ plus jets is simulated with up to five
additional partons at leading order.  Additional jets are generated
via parton showering, which, together with fragmentation and
hadronisation, is performed by {\tt
  HERWIG}~\cite{Corcella:2000bw,Corcella:2002jc}. The {\tt
  MLM}~\cite{Mangano:2006rw} prescription is used for matching the
matrix-element calculations to the parton shower evolution. {\tt
  JIMMY}~\cite{Butterworth:1996zw} is used to simulate the underlying
event. Additional $Z/W$ plus jets samples generated with {\tt
  SHERPA}~\cite{Gleisberg:2008ta} are used to estimate the
uncertainties related to the event generator. Single top quark and
pair production are simulated with {\tt
  MC@NLO}~\cite{Frixione:2006he}, fixing the top-quark mass to
$172.5\GeV$, and using the next-to-leading-order (NLO) PDF set {\tt
  CTEQ6.6}~\cite{Nadolsky:2008zw}. Parton showering and hadronisation
are performed with {\tt HERWIG}, and {\tt JIMMY} is again used for the
underlying event. Di-boson ($WW$, $WZ$, $ZZ$) samples are generated
with {\tt SHERPA}. Backgrounds from QCD multijet production are
estimated from data (see section~\ref{sec:bg:qcd} ). {\tt
  PYTHIA}~\cite{Sjostrand:2006za} simulations of this process,
normalised to data, are used in figures for illustrative purposes
only.

For graviton production in the ADD model, a low-energy effective field
theory~\cite{Giudice:1998ck} with energy scale $M_\mathrm{D}$ is used
to calculate the signal cross section considering the contribution of
different graviton mass modes. Signal samples corresponding to a
number of extra dimensions varying between two and six are considered,
with the renormalisation and factorisation scales set per event to
$\sqrt{\frac{1}{2} M_\mathrm{G}^2 + \pt^2}$, where $M_\mathrm{G}$ is
the mass of the graviton mode produced in the event and \pt\ denotes
the transverse momentum of the recoiling parton. The samples are
produced with an ADD implementation as a user model of {\tt PYTHIA},
which is also used for parton showering and hadronisation. {\tt
  MSTW2008\,LO$^{**}$}~\cite{Martin:2009iq} PDF sets are used for the
event simulation. The event yields for {\tt CTEQ6.6} PDFs are obtained
by re-weighting these samples, and are used to estimate cross
sections, as well as PDF systematic uncertainties. ADD cross sections
are calculated at both leading order (LO) and NLO. The NLO
calculations take into account QCD corrections to graviton production
and have been produced for the kinematic regions explored here
following ref.~\cite{Karg:2009xk}.

The effective field theory of WIMP pair production is implemented in
{\tt MADGRAPH5}~\cite{Alwall:2011uj} (version $1.3.33$), taken from
ref.~\cite{Goodman:2010ku}. WIMP pair production plus one and two
additional partons from ISR/FSR is simulated requiring at least one
parton with a minimum transverse momentum of $80\GeV$. Only initial
states of gluons and the four lightest quarks are considered, assuming
equal coupling strengths for all quark flavours to the WIMPs. The mass
of charm quarks is most relevant for the cross sections of the
operator D1 (see table~\ref{table:wimp:operators}) and it is set to
$1.42\GeV$. The generated events are interfaced to {\tt PYTHIA} for
parton showering and hadronisation. The {\tt MLM} prescription is used
for matching the matrix-element calculations of {\tt MADGRAPH5} to the
parton shower evolution of {\tt PYTHIA}. The {\tt CTEQ6L1} PDF set is
used for the event simulation. The {\tt MADGRAPH5} default choice for
the renormalisation and factorisation scales is used. The scales are
set to the sum of $\sqrt{m^2 + \pt^2}$ for all produced particles,
where $m$ is the mass of particles. Events with WIMP masses between 10
and $1300\GeV$ are simulated for four different effective operators
(D1, D5, D9, D11). In all cases, WIMPs are taken to be Dirac fermions,
and the pair-production cross section is calculated at LO.
 
The background MC samples use a detector simulation~\cite{Aad:2010wqa}
based on {\tt GEANT4}~\cite{GEANT4} and are reconstructed with the
same algorithms as the data. The signal MC samples employ a mix of the
detailed {\tt GEANT4} detector simulation and a simulation relying on
parametrisations of calorimetric signals to shorten the CPU time
required ({\tt ATLFAST-II}~\cite{Aad:2010wqa}).  Individual signal MC
samples have been validated against the more detailed detector
simulation relying fully on {\tt GEANT4}. Effects of \emph{event
  pile-up}---multiple $pp$ interactions occurring in the same or
neighbouring crossing of two proton bunches, called \emph{pile-up}
from now on---are included in the simulation. MC events are
re-weighted to reproduce the distribution of the number of collisions
per bunch crossing observed in the data.

\section{Analysis strategy and physics object reconstruction}
\label{sec:object:reco}
A search for a BSM excess is performed in monojet final
states. Leptons are vetoed to suppress background contributions from
$Z/W$ plus jets. A second jet is allowed as long as it is not aligned
with \met, which would be the case in multijet background events with
a mis-measured jet. Events with more than two jets are vetoed.

These selection requirements define the basic \emph{signal region}
(SR), defined in detail in table~\ref{tab:SRCR} below. The data sample
consisting of events passing the SR selections is sub-divided into
overlapping kinematic regions by applying selection criteria on \met\
and $\pt^\mathrm{jet1}$, the transverse momentum of the most energetic
jet (leading jet) in the event. Events in these overlapping individual
signal regions are then used to search for a BSM excess above the
predicted SM backgrounds.

The main SM background contributions to the SR data samples are from
$Z/W$+jets production and they are estimated from data by selecting
events based on a set of selection requirements---orthogonal to those
of the signal regions---that define a \emph{control region} (CR).
These CR requirements are based on selecting events with leptons
(either exactly one or two leptons). Different physics processes are
used to estimate background contributions to a SR: $W\ra e\nu$+jets,
$W\ra \mu\nu$+jets, $Z\ra\ee$+jets, and $Z\ra\mumu$+jets. To obtain
data samples enriched by these processes, a set of selection criteria
defines corresponding CR's. Each set of CR requirements is further
sub-divided into the same kinematic categories as the signal regions.
 
This analysis is based on reconstructed jets, electrons, muons, and
\met. The definitions of electron and muon candidates and of \met\ are
different in the SR and CR requirements. All electron candidates are
required to have $\pt > 20\GeV$ and $|\eta| < 2.47$, in order to be
within the acceptance of the tracking system. For the signal-region
requirements, which comprise an electron veto, relatively loose
criteria are used to define an electron candidate
(\emph{SR-electron}), because a looser electron definition leads to a
more stringent veto. SR-electrons are required to pass the
\emph{medium} electron shower shape and track selection criteria
described in ref.~\cite{Aad:2011mk}. No spatial isolation is
required. For the control-region selection requirements, used for
background estimates from events with measured electrons, more
stringent electron selection criteria (defining the
\emph{CR-electron}) are used in case exactly one electron is
selected. This is to better suppress jet contamination in these
control regions. A CR-electron is required to pass the
\emph{tight}~\cite{Aad:2011mk} electron shower shape and track
selection criteria in $W\ra e\nu$ control regions. In addition, the
following isolation criterion is imposed for CR-electrons to suppress
events where a jet is mis-identified as an electron: the scalar sum of
the transverse momentum of tracks with $\Delta R \equiv
\sqrt{(\Delta\phi)^2 + (\Delta\eta)^2} < 0.2$ around the electron
candidate, excluding the electron itself, has to be less than 10\% of
the electron's transverse energy $(E_\mathrm{T})$. In control regions
where exactly two electrons are required, the looser SR-electron
definition without isolation requirements is used.

A muon candidate used in the definition of the signal regions
(\emph{SR-muon}) is reconstructed either by associating a stand-alone
muon spectrometer track with an inner detector track, or from an inner
detector track that is confirmed by a directional segment in the muon
spectrometer~\cite{Aad:2010yt}. SR-muons, which are used as veto in
signal-region selections, are required to have $\pt > 7\GeV$ and
$|\eta| < 2.5$. They are also required to be isolated: the scalar \pt\
sum of tracks within $\Delta R = 0.2$ around the muon track, excluding
the muon itself, must be less than $1.8\GeV$. As for electrons, the
muon selection criteria in control-regions definitions are more
stringent. A \emph{CR-muon} candidate must have a stand-alone muon
spectrometer track associated with an inner detector track. Those
SR-muons that have only an inner detector track tagged by a segment in
the muon spectrometer do not satisfy the CR selection
criteria. Furthermore, CR-muons satisfy $\pt > 20\GeV$ and $|\eta| <
2.4$, and have an impact parameter along $z$ with respect to the
reconstructed primary vertex of $|z_0| < 10$~mm to reject cosmic-ray
muons. The CR-muons are also required to be isolated: the scalar \pt\
sum of tracks within $\Delta R = 0.2$ around the muon track, excluding
the muon itself, must be less than 10\% of the muon \pt.

In the signal regions, the measurement of \met\ is performed using all
clusters of energy deposits in the calorimeter up to $|\eta|$ of 4.5.
The calibration of these clusters takes into account the different
response of the calorimeters to hadrons compared to electrons or
photons, as well as dead material and out-of-cluster energy
losses~\cite{met_loc_had,Aad:2012re}. In the control regions, two
additional definitions of \met\ are used to account for the different
treatment in the signal and control regions of electrons and
muons. This is because the calorimetric definition of the nominal
\met\ takes into account energy deposits of electrons whereas it does
not account for transverse momentum carried away by muons. The two
additional definitions of \met\ either exclude the electron
contributions to the missing transverse momentum in events with
electrons or include the muon contributions to the missing transverse
momentum in events with muons:
\begin{itemize}
\item $E_{\mathrm{T}}^{\mathrm{miss}, \not e}:$ for control regions
  that involve electrons ($W\ra e\nu$+jets, $Z\ra \ee$+jets, explained
  in more detail below), $E_{\mathrm{T}}^{\mathrm{miss}, \not e}$ is
  obtained by adding the electron clusters to the missing transverse
  momentum vector thereby removing the electron contribution to the
  calculation of \met: $E_{\mathrm{T}}^{\mathrm{miss},
    \not e} = |{\bf p}^{\mathrm{miss}}_{\mathrm{T}} + {\bf
    p}_{\mathrm{T}}^{\mathrm{electrons}}|$. This yields missing
  transverse momentum which, as in invisible $Z$ decays, does not take
  into account the decay products of the $Z$ boson.
\item $E_{\mathrm{T}}^{\mathrm{miss},\mu}:$ the second alternative
  version of \met\ takes into account the muon contribution to \met\
  and it is used in the exclusive $W\ra \mu\nu$+jets control
  regions. It is defined as the negative sum of the calorimeter-based
  ${\bf p}^{\mathrm{miss}}_{\mathrm{T}}$ and the transverse momentum
  of muons, which do not deposit much energy in the calorimeters:
  $E_{\mathrm{T}}^{\mathrm{miss},\mu} = |{\bf
    p}^{\mathrm{miss}}_{\mathrm{T}} - {\bf
    p}_{\mathrm{T}}^\mathrm{muons}|$.
\end{itemize}
With these three versions of missing transverse momentum, the
kinematics of invisible $Z\ra \nu\bar{\nu}$ decays can be mimicked in
$Z$ or $W$ events with measured muons (\met) or electrons
($E_{\mathrm{T}}^{\mathrm{miss}, \not e}$). On the other hand, for the
selection of such control samples enriched with $Z$ or $W$ events, the
missing transverse momentum taking into account all visible decay
products of $Z$ or $W$ bosons can be used in events with measured
muons ($E_{\mathrm{T}}^{\mathrm{miss},\mu}$) or electrons (\met).

Jet candidates are reconstructed using the anti-$k_t$ clustering
algorithm~\cite{paper:antikt} with a radius parameter of $0.4$. The
inputs to this algorithm are clusters of energy deposits in
calorimeter cells seeded by those with energies significantly above
the measured noise~\cite{clusters}. Jet momenta are calculated by
performing a four-vector sum over these cell energy clusters, treating
each cluster as an $(E,{\bf p})$ four-vector with zero mass. The
direction of ${\bf p}$ is given by the line joining the nominal
interaction point with the calorimeter cluster. The resulting jet
energies are corrected to the hadronic scale using $\pt$ and $\eta$
dependent calibration factors based on MC simulations and validated by
extensive test beam and collision data studies~\cite{Aad:2011he}.

\section{Event selection}
\label{sec:evsel}
All data passing detector quality requirements are considered for the
analysis. Events must be accepted by an inclusive \met\
trigger~\cite{ATL-DAQ-PUB-2011-001,ATLAS-CONF-2011-072} that is found
to be 98\% efficient for events with \met\ above $120\GeV$, and more
than 99\% for \met\ above $150\GeV$. At $120\GeV$, a small residual
dependence on pile-up of the \met\ trigger efficiency is found. Over
the full 2011 dataset, where the pile-up varied from an average of 3
interactions per bunch crossing at the beginning of the year to 17 at
the end of the year, an efficiency variation of 1.5\% is observed and
a correction is applied to account for this variation. For \met\ above
$220\GeV$, there is no measurable efficiency variation.  Events are
further required to satisfy a set of pre-selection and kinematic
criteria that are aimed at selecting monojet events from good-quality
$pp$ collisions, as well as reducing electroweak, multijet,
non-collision, and detector-induced backgrounds. These criteria
require the event to have a monojet topology characterised by one
unbalanced high-\pt\ jet resulting in large $\met$:
\begin{itemize}
\item A reconstructed primary vertex with at least two associated
  tracks (with $\pt > 0.4\GeV$) is
  required~\cite{ATLAS-CONF-2012-042}.  This ensures that the recorded
  event is consistent with a proton-proton collision rather than a
  noise event.
\item The highest-\pt\ jet must have a charge fraction $f_\textrm{ch}=\sum
  \pt^\textrm{track,jet} /\pt^\textrm{jet} > 0.02$, where $\sum
  \pt^\textrm{track,jet}$ is the scalar sum of the transverse momenta
  of tracks associated with the primary vertex within a cone of radius
  $\Delta R=0.4$ around the jet axis, and $\pt^\textrm{jet}$ is the
  transverse momentum of the jet as determined from calorimeter
  measurements.  Furthermore, events are rejected if they contain any
  jet with an electromagnetic fraction $f_\textrm{em}$ (fraction of
  the jet energy measured in the electromagnetic calorimeter) of less
  than $0.1$, or any jet in the pseudorapidity range $|\eta|<2$ with
  $f_\textrm{em} >0.95$ and a charge fraction $f_\textrm{ch} \le
  0.05$.  These requirements suppress jets produced by cosmic rays or
  beam-background muons that interact in the hadronic calorimeter
  without corresponding signals in the electromagnetic calorimeter or
  the tracking detector.
\item Additional selection criteria to reject events with significant
  detector noise and non-collision backgrounds are applied: events are
  rejected if any jet with $\pt > 20\GeV$ and $|\eta| < 4.5$ does not
  pass all of the additional quality criteria described in
  ref.~\cite{atlas-jet-clean}.
\item The leading jet has to be within $|\eta|<2$, and no more than
  two jets with $\pt > 30\GeV$ and $|\eta|<4.5$ are
  allowed. Back-to-back dijet events are suppressed by requiring the
  sub-leading jet not to point in the direction of \ptmiss:
  $|\Delta\phi(\ptmiss,\jettwo)|>0.5$.
\item An electronics failure affecting 20\% of the data sample created
  a small dead region in the second and third layers of the
  electromagnetic calorimeter. Any event with the two leading jets
  inside the affected region and either of the two jets pointing in
  the direction of \met\ is removed from the sample to avoid fake
  signals. This condition removes only a few percent of the affected
  subset of the data.
\item Events are required to have no SR-electron or SR-muon. In the
  background control regions, electrons and muons are explicitly
  selected.  The electron and muon selection criteria in the signal
  and control regions are given in section~\ref{sec:object:reco}.
\end{itemize}

Although the results of this analysis are interpreted in terms of the
ADD model and WIMP pair production, the event selection criteria have
not been tuned to maximise the sensitivity to any particular BSM
scenario. To maintain sensitivity to a wide range of BSM models, four
sets of overlapping kinematic selection criteria, designated as SR1 to
SR4, differing in the values of the requirements for \met\ and leading
jet \pt, are defined~(table~\ref{tab:SRCR}). Note that the requirement
on the leading jet \pt\ is the same as that on \met\ for all signal
regions. In comparison with the previous ATLAS monojet
search~\cite{Aad:2011xw}, the veto on additional jets is less
stringent, allowing a second jet in the event thereby reducing
systematic uncertainties from ISR/FSR (see section~\ref{sec:results})
and increasing signal selection efficiencies. The signal region with
the lowest \met\ requirement (SR1) is chosen such that the $\met$
trigger is nearly 100\% efficient. The signal region with the highest
\met\ requirement (SR4) is chosen so that there remain enough events
in data control samples to validate MC predictions and estimate SR
backgrounds in a data-driven way.
\begin{table}[tbp]
  \begin{center}
    \begin{tabular}{|c|cccc|}
  \hline
  Signal regions  & SR1 & SR2 & SR3 & SR4 \\ \hline
  &\multicolumn{4}{c|}{Data quality + trigger + vertex + jet quality +}\\
  Common requirements & \multicolumn{4}{c|}{$|\eta^{\mathrm{jet1}}| < 2.0$ +
    $|\Delta\phi(\ptmiss,\jettwo)|>0.5$ + $N_\mathrm{jets}\le 2$ +} \\ 
  & \multicolumn{4}{c|}{lepton veto} \\\hline
  $\met$, $\pt^\mathrm{jet1} >$ & $120\GeV$ 
  & $220\GeV$ 
  & $350\GeV$ 
  & $500\GeV$ \\
  \hline
\end{tabular}
\caption{\label{tab:SRCR} Definition of the four overlapping signal
  regions SR1--SR4. \emph{Data quality}, \emph{trigger},
  \emph{vertex}, and \emph{jet quality} refer to the selection
  criteria discussed in the main text.}
\end{center}
\end{table}

\section{Background estimation}
\label{sec:bg}
A number of SM processes can pass the monojet kinematic selection
criteria described above.  These backgrounds include, in decreasing
order of importance: $Z$ and $W$ boson plus jets production, single or
pair production of top quarks, multijet production, cosmic-ray and
beam-background muons\footnote{Originating either from protons going
  in the direction of the experiment and hitting the LHC collimation
  system or gas molecules in the beam-pipe near the ATLAS interaction
  point.} (collectively referred to as \emph{non-collision
  background}~\cite{ATLAS-CONF-2011-137}), and di-boson production
($WW$, $WZ$, $ZZ$).  The dominant $Z/W$ plus jets backgrounds are
estimated using control regions in the data with corrections that
account for differences between the selection requirements of the
signal and control regions~(see section~\ref{sec:bg:1}). The multijet
and non-collision backgrounds are also estimated from data~(see
sections~\ref{sec:bg:qcd} and \ref{sec:bg:ncb}, respectively) while
the di-boson and top-quark backgrounds are obtained from MC
simulations.

\subsection{Backgrounds from \emph{Z/W}+jets}
\label{sec:bg:1}
The dominant background process for this search is irreducible and
consists of the production of $Z$ bosons in association with jets,
where the \Zboson\ decays to two neutrinos.  A substantial source of
reducible background is SM \Wboson\ boson plus jets production where
the \Wboson\ decays to a charged lepton ($\tau$, $e$, or $\mu$ in
decreasing order of importance) and a neutrino. This process leads to
a monojet final state if the lepton is outside the detector
acceptance, is missed because of reconstruction inefficiencies or if a
hadronic $\tau$ decay is reconstructed as a single jet. The \Zboson\
and \Wboson\ boson plus jets backgrounds, collectively referred to in
the following as \emph{electroweak backgrounds}, are determined in a
data-driven way:
\begin{enumerate}
\item Control regions are defined by explicitly selecting electrons or
  muons while keeping the same jet and \met\ selection criteria as in
  the signal regions. In a first step, samples enriched with four
  processes containing electrons or muons are separately selected with
  dedicated selection requirements: $W\ra e\nu$+jets, $W\ra
  \mu\nu$+jets, $Z\ra\ee$+jets, $Z\ra\mumu$+jets. In a second step,
  the jet and \met\ selection criteria as in the signal regions are
  imposed. Corrections are made for contamination of these control
  samples from processes other than \Zboson\ or \Wboson\ decays.
\item Correction factors are then applied to account for differences
  in trigger and kinematic selection criteria between the control and
  signal regions.  The control-to-signal region transfer factors,
  which are multiplied by the number of control-region events obtained
  in the previous step to yield the background estimate, are obtained
  using both data and simulation (see below).
\end{enumerate}
In this approach, the modelling of the jet and \met\ kinematics of the
electroweak backgrounds is obtained directly from data. Simulations
are therefore used only for quantities related to the electron and
muon selection criteria, and only through ratios where systematic
uncertainties related to the jet and \met\ selection criteria of the control
regions cancel. Theoretical uncertainties normally associated with MC
estimates are significantly reduced; only distributions related to the
electron and muon selection criteria have to be well modelled in the
simulations. Further experimental uncertainties that impact the
background prediction, such as the jet energy scale (JES) and
resolution (JER) ~\cite{Aad:2011he}, the trigger efficiency, and
the luminosity measurement~\cite{Aad:2011dr,confnote:lumi}, are minimised
by this approach.

\begin{table}[tbp]
\begin{center}
\begin{tabular}{|l|llll|}
  \hline
  \multirow{2}{*}{SR process} 
  & \multirow{2}{*}{$Z\ra \nu\bar{\nu}$+jets}
  &$W\ra \tau\nu$+jets
  &\multirow{2}{*}{$W\ra e\nu$+jets}
  &$Z\ra \tautau$+jets\\
  &
  &$W\ra \mu\nu$+jets
  &
  &$Z\ra \mumu$+jets\\\hline
  \multirow{4}{*}{CR process} 
  &$W\ra e\nu$+jets
  &\multirow{4}{*}{$W\ra \mu\nu$+jets}
  &\multirow{4}{*}{$W\ra e\nu$+jets}
  &\multirow{4}{*}{$Z\ra \mumu$+jets}\\
  &$W\ra \mu\nu$+jets
  &
  &
  &\\
  &$Z\ra \ee$+jets&&&\\
  &$Z\ra \mumu$+jets&&&\\\hline
\end{tabular}
\caption{\label{tab:bg:2}Overview of processes in the control regions
  (CR) used to estimate background contributions to processes in the
  signal regions (SR).}
\end{center}
\end{table} 
The control regions are expected to have no contamination from BSM
signals that would normally pass the monojet event selection
criteria. They are chosen such that they are dominated by $Z$ and $W$
decays with reconstructed electrons or muons. The selection criteria
follow closely those used in $Z$ and $W$ cross-section
measurements~\cite{Aad:2011dm}.  The kinematic selection criteria on
\met\ and jet \pt\ of the signal regions are also applied. Therefore,
each CR is split into four subsets corresponding to the four signal
regions. Four visible decay modes are used for the background
estimates: $W\ra e\nu$+jets, $W\ra \mu\nu$+jets, $Z\ra\ee$+jets,
$Z\ra\mumu$+jets. Based on these, all contributions to the signal
regions from $Z$ and $W$ decay modes are estimated with the same
method (except for $\Zboson\ \ra \ee$+jets, which is found to be
negligible in the signal regions because both $e^+$ and $e^-$ would
have to be missed in the event selection). In total, six background
processes in each signal region are predicted based on four control
region processes, as detailed in table~\ref{tab:bg:2}. Note that the
$W\ra \tau\nu$+jets background in the signal regions, where the $\tau$
lepton decays hadronically, can safely be estimated from $W\ra
\mu\nu$+jets in control regions, since the jet and \met\ kinematics
are the same. In both cases the leading jet is from radiation and
recoils against the neutrino from the \Wboson\ decay. The hadronic
$\tau$ decay results in a jet that is either below the jet threshold
of $30\GeV$ or above this threshold but still sub-dominant compared to
the leading jet from radiation.

For control regions that include processes with electrons, an electron
trigger is used that requires a correction to account for differences
in efficiency and acceptance compared to the \met\ trigger used for
the signal regions.\footnote{Note that the \emph{acceptance} is
  defined as the ratio of the number of events within the detector
  volume that pass analysis requirements to the number of originally
  simulated events. The \emph{efficiency} is defined as the ratio of
  the number of events within the detector volume at reconstruction
  level to that at the original simulation level.} This different
treatment is required because the energy deposited by electrons is
included in the \met\ measurement at trigger level and results in the
selection of a different kinematic region than that of the signal
regions, which exclude electrons. Muons, however, do not deposit large
amounts of energy in the calorimeters and are not explicitly included
in the \met\ trigger. The specific selection criteria for the four
control region processes are given in the following:
\begin{itemize}
\item \underline{$W\ra e\nu$+jets:} Events are selected using electron
  triggers with thresholds of 20 or $22\GeV$ depending on the
  data-taking period. The CR-electron definition is used (see
  section~\ref{sec:object:reco}) and exactly one electron with a \pt\
  of at least $25\GeV$ is required. Events with additional electrons
  or muons are discarded. All triggers used are fully efficient above
  the chosen \pt\ cut value. If an object is reconstructed as both an
  electron and a jet, the jet is removed from the reconstructed jet
  collection if $\Delta R (e,\mathrm{jet}) < 0.2$ while the electron
  is kept. To further improve the $W$ purity, $\met > 25\GeV$ and $40
  < m_\mathrm{T} < 100\GeV$ are required. $m_\mathrm{T}$ is the
  transverse mass and it is defined as $m_\mathrm{T} = \sqrt{2 \, \pt
    \, \met \, (1 -
    \cos{\Delta\phi(\mathbf{p}_{\mathrm{T}}^\mathrm{lepton},\ptmiss)})}$,
  using the \pt\ of the lepton (electron or muon). $\Delta\phi$ is the
  angle between the lepton and the missing transverse momentum
  vector. As mentioned earlier, the same selection criteria on jet
  \pt\ and \met\ are applied in the control regions as in the signal
  regions (see table~\ref{tab:SRCR}).  However, when the $W\ra
  e\nu$+jets CR is used to estimate the contribution of $Z\ra
  \nu\bar{\nu}$+jets to each SR, a special CR is defined where \met\
  is substituted by $E_{\mathrm{T}}^{\mathrm{miss}, \not e}$ to mimic
  the kinematics of the decay of the $Z$ boson to two undetected
  neutrinos. The standard calorimeter-based \met\ is used for the CR
  to estimate the $W\ra e\nu$+jets contribution to the SRs.
\item \underline{$W\ra \mu\nu$+jets:} Events have to pass the same
  inclusive \met\ trigger that is used for the signal regions. Exactly
  one CR-muon (see section~\ref{sec:object:reco}) is required, and
  events with additional electrons or muons are rejected. Cuts on the
  transverse mass and missing transverse momentum are applied to
  improve the purity for $W$'s: $m_\mathrm{T} > 40\GeV$,
  $E_{\mathrm{T}}^{\mathrm{miss},\mu} > 25\GeV$. Note that the \met\
  that includes the muon contribution,
  $E_{\mathrm{T}}^{\mathrm{miss},\mu}$, is used for the $W$-specific
  selection cuts. For each kinematic region listed in
  table~\ref{tab:SRCR}, the standard calorimeter-based \met\ is used
  to define the CRs for the estimates of both $Z\ra \nu\bar{\nu}$+jets
  and $W\ra \mu\nu$+jets in the corresponding SR.
\item \underline{$Z\ra \ee$+jets:} Electron triggers are used in this
  channel. Exactly two opposite-sign electrons are required and events
  with additional electrons or muons are discarded. The selected
  electrons have to satisfy $\pt > 25~(20)\GeV$ for the leading
  (sub-leading) electron. Jet-electron overlap removal is performed as
  described above for $W\ra e\nu$+jets. Finally, to enhance the
  fraction of $Z$'s, an invariant mass requirement of $66 < m_{\ee} <
  116\GeV$ is applied. $E_{\mathrm{T}}^{\mathrm{miss}, \not e}$ is
  used to define these CRs, which are used to estimate the $Z\ra
  \nu\bar{\nu}$+jets contribution to the SRs.
\item \underline{$Z\ra \mumu$+jets:} The inclusive \met\ trigger is
  used in this channel. Exactly two opposite-sign CR-muons (defined in
  section~\ref{sec:object:reco}) are required and events with
  additional electrons or muons are rejected. An invariant mass of $66
  < m_{\mumu} < 116\GeV$ is required to select $Z$ candidates. The
  signal-region selection criteria are then applied on the calorimeter-based
  \met, for the $Z\ra \nu\bar{\nu}$+jets, $Z\ra \tautau$+jets, $Z\ra
  \mumu$+jets estimates.
\end{itemize}
Using the control regions defined above, the background contribution
to the signal regions for each combination of CR and SR processes
mentioned in Table~\ref{tab:bg:2} is derived using:
\begin{eqnarray}
  &&N^\mathrm{predicted}_\mathrm{SR} = (N^\mathrm{Data}_\mathrm{CR} - N_\mathrm{CR}^\mathrm{Bkg}) \cdot C \cdot
  T = \nonumber\\[0.4cm]
  &&\left((N^\mathrm{Data}_\mathrm{CR}-N_\mathrm{CR}^\mathrm{multijet})
  \cdot (1 - f_\mathrm{EW})\right) \times \frac{\epsilon^\mathrm{trig}_{\met} \cdot \mathcal{L}_{\met}}{A_\mathrm{\ell} \cdot
    \epsilon_\mathrm{\ell} \cdot \epsilon_{Z/W} \cdot \epsilon^\mathrm{trig}_\mathrm{\ell} \cdot
    \EuScript{L}_\mathrm{\ell}} \times \frac{N^\mathrm{MC}_\mathrm{SR}}{N^\mathrm{MC}_\mathrm{jet/\met}} \,
  . \label{eq:bg:2}
\end{eqnarray}
Data in the control regions are corrected for contamination arising
from other sources (summarised as $N_\mathrm{CR}^\mathrm{Bkg}$ in the
first line). Correction factors $(C)$ based on MC simulation
and data are applied together with the transfer factor $T$ to obtain
the number of background events, $N^\mathrm{predicted}_\mathrm{SR}$,
predicted in the signal region. The terms appearing in the second line
of equation~\ref{eq:bg:2} are:
\begin{itemize}
\item $N^\mathrm{Data}_\mathrm{CR}$ and
  $N_\mathrm{CR}^\mathrm{multijet}$ are the number of data and
  multijet events in the control region, respectively.  To estimate
  the multijet contamination of control regions by processes with
  identified electrons, the selection cuts, in particular the
  isolation cuts, are varied. The fake rate in those regions is
  extracted from data using real and fake electron efficiencies
  determined from samples enriched in electrons and jets.  Using this
  estimate, the multijet contamination is predicted to account for
  1-2\% of the events in the $W\ra e\nu$+jets electron control region
  when predicting the $Z\ra \nu\bar{\nu}$+jets contribution to the
  SRs. For other control regions containing electrons or muons, the
  multijet contamination is found to be negligible using similar
  techniques.
\item $f_\mathrm{EW}$ is the estimated fraction of events, after
  multijet corrections, due to contamination of the control region by
  other electroweak or other SM processes. This contamination is due
  to top-quark and di-boson decays as well as decays of $Z$ or $W$
  bosons to leptons of a flavour other than the one selected for that
  control region. The contribution of this contamination is obtained
  from MC simulation and is about 2\% for $Z$ bosons, and 10\% for $W$
  bosons. The top-quark and di-boson contribution is negligible. As
  explained above, using ratios of MC estimates ($f_\mathrm{EW}$ in
  this case) is advantageous as it leads to cancellations of
  systematic uncertainties.
\item $A_\mathrm{\ell}$ and $\epsilon_\mathrm{\ell}$ are the lepton
  acceptance obtained from simulation and the identification
  efficiency obtained from data~\cite{Aad:2010yt,Aad:2011mk},
  respectively.
\item $\epsilon_{Z/W}$ are the efficiencies for the $Z$ or $W$ boson
  selection criteria obtained from simulation. The factors
  $A_\mathrm{\ell}$, $\epsilon_\mathrm{\ell}$, and $\epsilon_{Z/W}$
  correct for the fact that leptons and $Z$/$W$ bosons are required
  only in the control regions.
\item $\epsilon_\mathrm{\ell}^\mathrm{trig}$ and
  $\EuScript{L}_\mathrm{\ell}$ are the electron trigger efficiency
  (obtained from data) and the corresponding luminosity associated
  with this trigger for the relevant control region.  For muon control
  regions these factors do not apply because the signal-region trigger
  is used in the definition of the CR.
\item $\epsilon_{\met}^\mathrm{trig}$ and $\EuScript{L}_{\met}$ are
  the $\met$ trigger efficiency (obtained from data) and the
  corresponding luminosity, and are only relevant for electron control
  regions where the electron trigger efficiency and luminosity
  ($\epsilon_{\ell}^\mathrm{trig}$ and $\EuScript{L}_\mathrm{\ell}$)
  need to be corrected, accounting for the different triggers used in
  the definition of the signal and control regions.
\item The transfer factor $T =
  \frac{N^\mathrm{MC}_\mathrm{SR}}{N^\mathrm{MC}_\mathrm{jet/\met}}$
  is the ratio of simulated background events in the signal region
  (for example $Z\ra \nu\bar{\nu}$+jets) to simulated events of a
  control-region process (for example $W\ra e\nu$+jets) with only jet
  and \met\ related selection requirements applied. This term
  translates the number of observed events in the CR in the data to
  the predicted number of events in the signal region. Depending on
  the control region and the signal-region background component being
  determined, this factor can account for ratios of branching
  fractions, ratios of $W$+jets to $Z$+jets cross sections, and
  phase-space differences between the control and signal regions for a
  given source of background.
\end{itemize}
The correction factors and electroweak background predictions in
equation~\ref{eq:bg:2} are determined in bins of \met\ for the final
background prediction, and in bins of the leading and sub-leading jet
\pt\ for the jet-\pt\ plots in figure~\ref{fig:sr1sr4}.

\subsection{Multijet backgrounds}
\label{sec:bg:qcd}
Multijet events where one or more jets are severely mismeasured
constitute a background that is not well modelled in the simulation.
In order to measure this background from data, a sample is selected by
applying all signal-region selection criteria except for the jet
vetoes: A) either a second jet with
$|\Delta\phi(\ptmiss,\jettwo)|<0.5$ is required, B) or the third-jet
veto is reversed by requiring three jets, $N_\mathrm{jet} = 3$, and
missing transverse momentum to be aligned with the third jet:
$|\Delta\phi(\ptmiss,\jetthree)|<0.5$ and
$|\Delta\phi(\ptmiss,\jettwo)|>0.5$. These two samples are used to
predict the multijet background from the resulting di- or trijet
events. Contributions to these event samples from top-quark, and $Z$
or $W$ production are subtracted. The MC simulation is used for the
top-quark contribution. For $Z$ and $W$, MC estimates normalised to
data are used for the subtraction. The multijet background is then
estimated by fitting a straight line to the second or third jet \pt\
distributions in events passing the two selection criteria (A) and B))
and then extrapolating the fit below a \pt\ of $30\GeV$. For this
value of the transverse momentum, the jets fall below the threshold
and can pass the monojet selection criteria. Note that the number of
trijet events where both sub-leading jets are mismeasured, and fall
below the jet threshold, is negligible compared to the case where
either the second or the third jet is lost. The resulting background
estimates are given in table~\ref{tab:bg}. They are at most 1\% of the
total background predicted for SR1--SR3, and are negligible for SR4.

\subsection{Non-collision backgrounds}
\label{sec:bg:ncb}
Non-collision backgrounds in the signal regions are estimated using a
dedicated algorithm that identifies beam-background muons that go
through the detector along the direction of the beams.  The algorithm
selects through-going muons based on timing information obtained from
the muon chambers in the forward regions. It combines this information
with calorimeter energy clusters by matching them in $\phi$. Unpaired
proton bunches, where the bunch from one of the proton beams is empty,
are used to determine identification efficiencies of the algorithm for
beam-background and cosmic-ray muons. These efficiencies
($\epsilon^\mathrm{non-coll.}_\mathrm{tag}$), typically 20--50\%, are
used together with the number of beam-background (\emph{halo})
candidates found in the signal regions, to predict the level of
non-collision background ($N_\mathrm{non-coll.} = N_\mathrm{halo} /
\epsilon^\mathrm{non-coll.}_\mathrm{tag}$). More details of this
background component are given in ref.~\cite{ATLAS-CONF-2011-137}. It
contributes mainly in SR1 and SR2 at less than 1\%. The predictions
are given in table~\ref{tab:bg}.

\subsection{Systematic uncertainties on background estimates}
\label{sec:bg:systematic}
The dominant systematic uncertainties associated with the electroweak
background estimates are on JES and \met, as well as theoretical
uncertainties on the shape of $W$ kinematic distributions and the
ratio of $Z$ and $W$ plus jets production cross sections. The latter
theoretical uncertainty is relevant because background predictions
from $W$ control regions are also used to estimate $Z\ra
\nu\bar{\nu}$+jets contributions to the signal region. Additional
systematic uncertainties are due to the muon momentum scale and
resolution, the data-driven scale factors to equalise lepton trigger
and reconstruction efficiencies in simulations and data, statistical
uncertainties associated with the limited size of MC samples, the
subtraction of the electroweak contamination ($f_\mathrm{EW}$), and,
in the electron control regions, the multijet contamination. The
uncertainties from pile-up variations are found to be negligible, as
are those from the electron energy scale, resolution, and JER.

The systematic uncertainties associated with the small multijet
background are estimated to be 100\%.  They are obtained by changing
the fit range for the \pt\ extrapolation and varying the scale factors
for the $Z$ and $W$ background prediction by 10\%. All variations are
within a factor of two of the central predictions.

Systematic uncertainties on the non-collision background are
10\%. This estimate corresponds to the average fraction of unpaired
proton bunches that are used to determine
$\epsilon^\mathrm{non-coll.}_\mathrm{tag}$, and that are close
(separated by 25 ns in time) to an unpaired bunch in the opposite
beam. Such configurations may lead to double counting in the
efficiency estimate, and their total contribution is hence
considered as an uncertainty.

The JES and JER uncertainties are evaluated using a combination of
data-driven and MC-based techniques~\cite{Aad:2011he}. These methods
take into account the variation of the uncertainty with jet \pt\ and
$\eta$, and the presence of nearby jets. The \met\ uncertainty is
derived from the JES and JER uncertainties by propagating the relative
jet-level variations to the calorimeter cluster based \met. Since
ratios are used to extrapolate from the control regions to the signal
regions, the effects of these uncertainties tend to cancel.

Theoretical uncertainties on $Z$ and $W$ production and the shape of
$W$ kinematic distributions are evaluated by comparing background
estimates using kinematic $Z/W$ distributions from different
generators ({\tt ALPGEN} and {\tt SHERPA}). Uncertainties on
$f_\mathrm{EW}$ are derived by comparing {\tt ALPGEN} to {\tt
  PYTHIA}~\cite{Aad:2011xn}, but also by taking into account JES and
lepton-scale uncertainties. The full difference is taken as systematic
uncertainty in all cases.

Systematic and statistical uncertainties on all background estimates
are given in table~\ref{tab:syst_uncert}. The contribution from lepton
scale factors is the quadratic sum of electron and muon
uncertainties. The uncertainties from di-boson, top-quark, multijet,
and non-collision backgrounds are summed in quadrature. A 20\%
uncertainty is assigned for the di-boson and top-quark MC-based
estimates. This value is dominated by the JES uncertainty (16\%), but
also takes into account uncertainties of the trigger efficiency,
luminosity measurement, and lepton identification uncertainties.
\begin{table}[tbp]
\begin{center}
\begin{tabular}{|l|llll|}\hline
Source & SR1 & SR2 & SR3 & SR4\\\hline
JES/JER/\met\ & 1.0 & 2.6 & 4.9 & 5.8\\
MC $Z/W$ modelling &2.9&2.9&2.9&3.0\\
MC statistical uncertainty &0.5&1.4&3.4&8.9\\
$1-f_\mathrm{EW}$ &1.0&1.0&0.7&0.7\\
Muon scale and resolution & 0.03 & 0.02 & 0.08 & 0.61\\
Lepton scale factors & 0.4 & 0.5 & 0.6 & 0.7\\ 
Multijet BG in electron CR & 0.1 & 0.1 & 0.3 & 0.6\\
Di-boson, top, multijet, non-collisions& 0.8 & 0.7 & 1.1 & 0.3\\
\hline
Total systematic uncertainty &3.4&4.4&6.8&11.1\\
Total data statistical uncertainty &0.5&1.7&4.3&11.8\\\hline
\end{tabular}
\caption{\label{tab:syst_uncert} Relative systematic uncertainties for
  all signal regions (in percent). Individual contributions are summed
  in quadrature to derive the total numbers. The MC statistical
  uncertainty is included in the total systematic uncertainty.}
\end{center}
\end{table}

\subsection{Background summary and additional checks}
\label{sec:bg:summary}
An overview of all backgrounds is given in table~\ref{tab:bg} (cf.\
table~\ref{tab:bg:2} for the definition of the control regions). The
final $Z\ra \nu\bar{\nu}$+jets predictions are estimated from a
combination of the predictions of the four control regions. The
combination is the error-weighted average calculated taking into
account correlations of uncertainties. The $Z\ra \nu\bar{\nu}$+jets
prediction is dominated by the $W$ control-region estimates and based
on the assumption that the ratio of $Z$+jets to $W$+jets cross
sections is well modelled in the simulation. This assumption is
supported by dedicated measurements~\cite{Aad:2011xn}, albeit for
smaller jet momenta than the ones used in SR2 to SR4. The theoretical
uncertainty on the ratio of $Z$ to $W$ cross sections is included in
the uncertainty derived from comparisons of different MC generators
discussed above.
\begin{table}[tbp]
\begin{center}
\scalebox{0.99}{%
\setlength{\tabcolsep}{6pt}
\begin{tabular}{|l|cccc|}
  \hline
  &SR1 & SR2 & SR3 & SR4 \\ \hline
  $Z\rightarrow\nu\bar{\nu}$+jets
  & 63000 $\pm$ 2100  
  & 5300  $\pm$ 280  
  & 500   $\pm$ 40   
  & 58    $\pm$ 9     
\\
  $W\rightarrow\tau\nu$+jets
  & 31400 $\pm$ 1000 
  & 1853  $\pm$ 81    
  & 133   $\pm$ 13    
  & 13    $\pm$ 3     
  \\
  $W\rightarrow e \nu$+jets
  & 14600 $\pm$ 500
  & 679   $\pm$ 43   
  & 40    $\pm$ 8    
  & 5     $\pm$ 2    
  \\
  $W\rightarrow\mu\nu$+jets
  & 11100 $\pm$ 600
  & 704   $\pm$ 60   
  & 55    $\pm$ 6    
  & 6     $\pm$ 1    
  \\
  $t\bar{t}$ + single $t$
  & 1240 $\pm$ 250
  & 57   $\pm$ 12   
  & 4    $\pm$ 1    
  & -                     
  \\
  Multijets
  & 1100 $\pm$ 900 
  & 64   $\pm$ 64   
  & $8^{+9}_{-8}$    
  & - 
  \\
  Non-coll. Background
  & 575 $\pm$ 83 
  & 25  $\pm$ 13  
  & -
  & - 
  \\ 
  $Z/\gamma^*\rightarrow\tau\tau$+jets
  & 421 $\pm$ 25  
  & 15  $\pm$ 2   
  & 2   $\pm$ 1   
  & -                   
  \\
  Di-bosons
  & 302 $\pm$ 61
  & 29  $\pm$ 5   
  & 5   $\pm$ 1   
  & 1   $\pm$ 1  
  \\ 
  $Z/\gamma^*\rightarrow\mu\mu$+jets
  & 204 $\pm$ 19
  & 8   $\pm$ 4   
  & -                   
  & -                   
  \\\hline
  Total Background
  & 124000 $\pm$ 4000
  & 8800   $\pm$ 400   
  & 750    $\pm$ 60    
  & 83     $\pm$ 14    
  \\
  Events in Data $(4.7\ifb)$
  & 124703
  & 8631
  & 785
  & 77 \\ \hline
  $\sigma^{\mathrm{obs}}_{\mathrm{vis}}$ at $90\%$ [ pb ] &1.63&0.13&0.026&0.0055\\[0.2cm]
  $\sigma^{\mathrm{exp}}_{\mathrm{vis}}$ at $90\%$ [ pb ] &1.54&0.15&0.020&0.0064\\[0.2cm]
  $\sigma^{\mathrm{obs}}_{\mathrm{vis}}$ at $95\%$ [ pb ] &1.92&0.17&0.030&0.0069\\[0.2cm]
  $\sigma^{\mathrm{exp}}_{\mathrm{vis}}$ at $95\%$ [ pb ] &1.82&0.18&0.024&0.0079\\
  \hline
\end{tabular}}
\end{center}
\caption{Overview of predicted SM background and observed 
  events in data for $4.7\ifb$ for each of the four signal regions. The total
  uncertainty quoted is the quadratic sum of statistical and
  systematic uncertainties. Observed and expected 90\% and 95\% CL
  upper limits on the non-SM contribution to all signal regions are
  also given in terms of limits on visible cross sections
  $(\sigma_{\mathrm{vis}}\equiv \sigma \times A \times \epsilon)$. The
  90\% CL upper limits are given to facilitate comparisons with other
  experiments.} 
\label{tab:bg}
\end{table} 

The electroweak background estimate, which relies on an exclusive $W$
or $Z$ selection in the control regions, is compared to two
alternative correlated methods. In the first of these, which was the
main method used in the previous ATLAS monojet
search~\cite{Aad:2011xw}, an inclusive control region is defined by
only inverting the lepton veto while keeping all other selection
criteria the same as in the signal regions. No additional $Z$- or
$W$-specific invariant or transverse mass selection criteria are
applied, thereby yielding a mixed control sample dominated by $W$ and
$Z$ bosons. The resulting background predictions are found to be
consistent with those of the default method. The second alternative
modifies the lepton definition in the control regions. Instead of
applying lepton selection criteria in control regions that are more
stringent than those of the signal regions, a modified exclusive
control region is defined. The selection criteria include less
stringent lepton definitions where the lepton veto cuts of the signal
region are simply inverted, and dedicated $Z$ or $W$ selection
criteria are used. These background predictions are also found to be
consistent with the default method.

Distributions from all four visible decay modes used to determine the
background in SR1 are shown in figure~\ref{fig:cr}. The distributions
are obtained by applying the exclusive $Z$ and $W$ selection criteria
plus SR1 kinematic cuts on \met\ and jets, as well as vetoes on
additional electrons or muons.  It should be noted that shape
differences in the \met\ distributions between data and MC are
irrelevant for an accurate background prediction in the signal
regions, because the \met\ distribution obtained from control-region
data is used directly to predict the backgrounds in the signal
regions.  Distributions of variables that are subject to MC-based
efficiency or acceptance corrections, namely those involving electrons
or muons, need to agree in shape between data and MC (see
figure~\ref{fig:cr}, where good shape agreement is found for the
leading electron and muon \pt\ distributions).
\begin{figure}[htb]
\begin{center}
\includegraphics[width=0.49\linewidth]{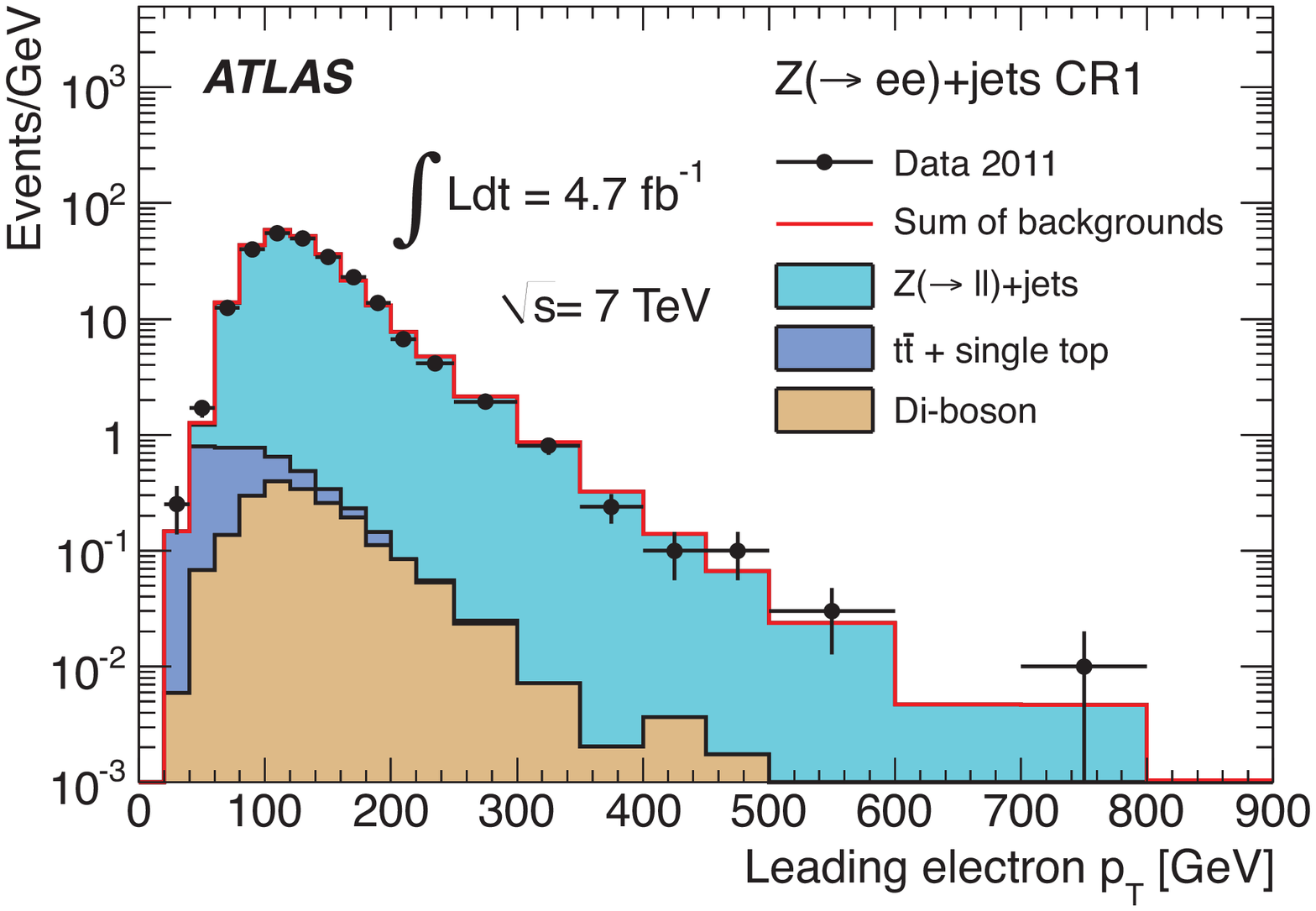}
\includegraphics[width=0.49\linewidth]{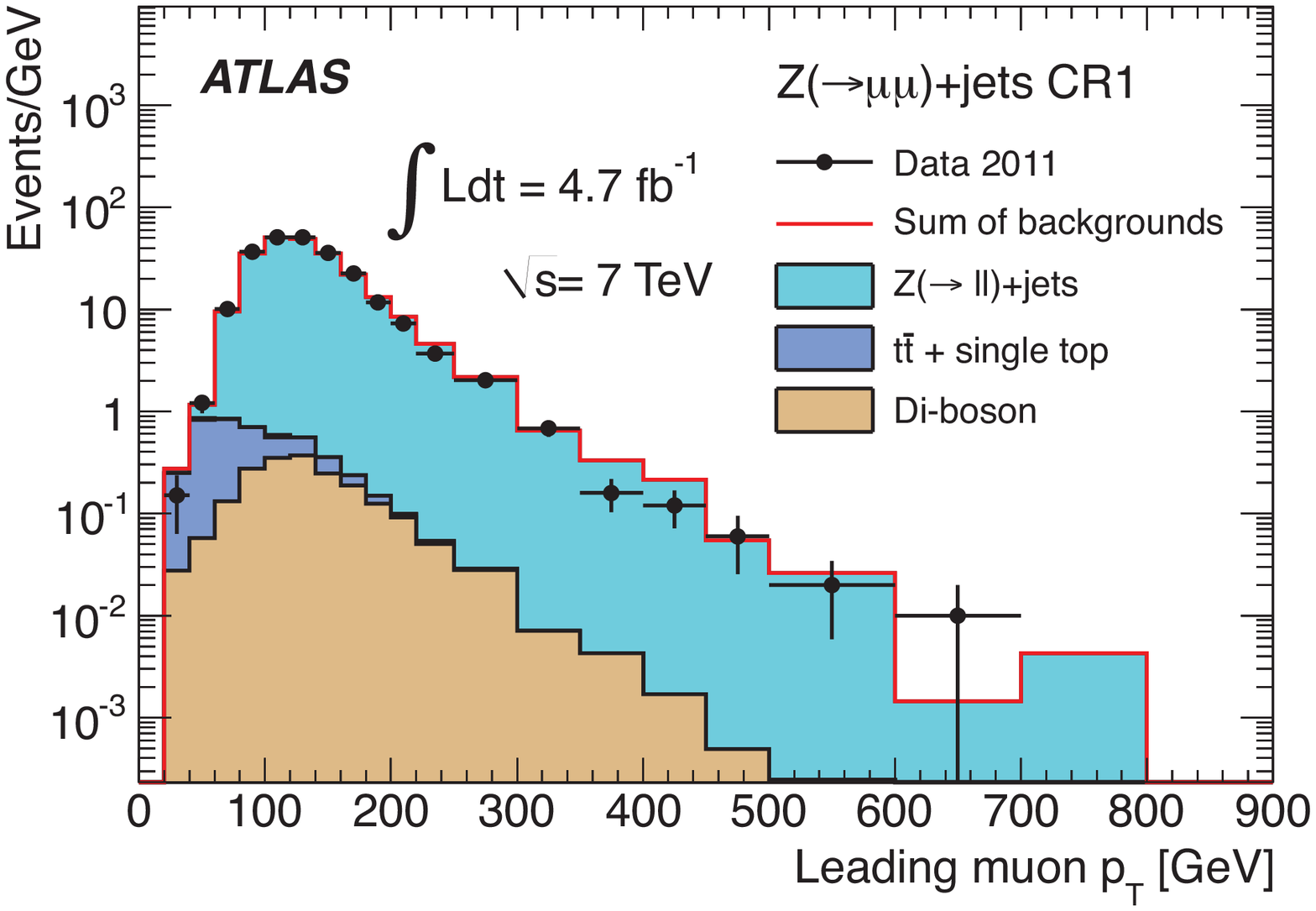}
\includegraphics[width=0.49\linewidth]{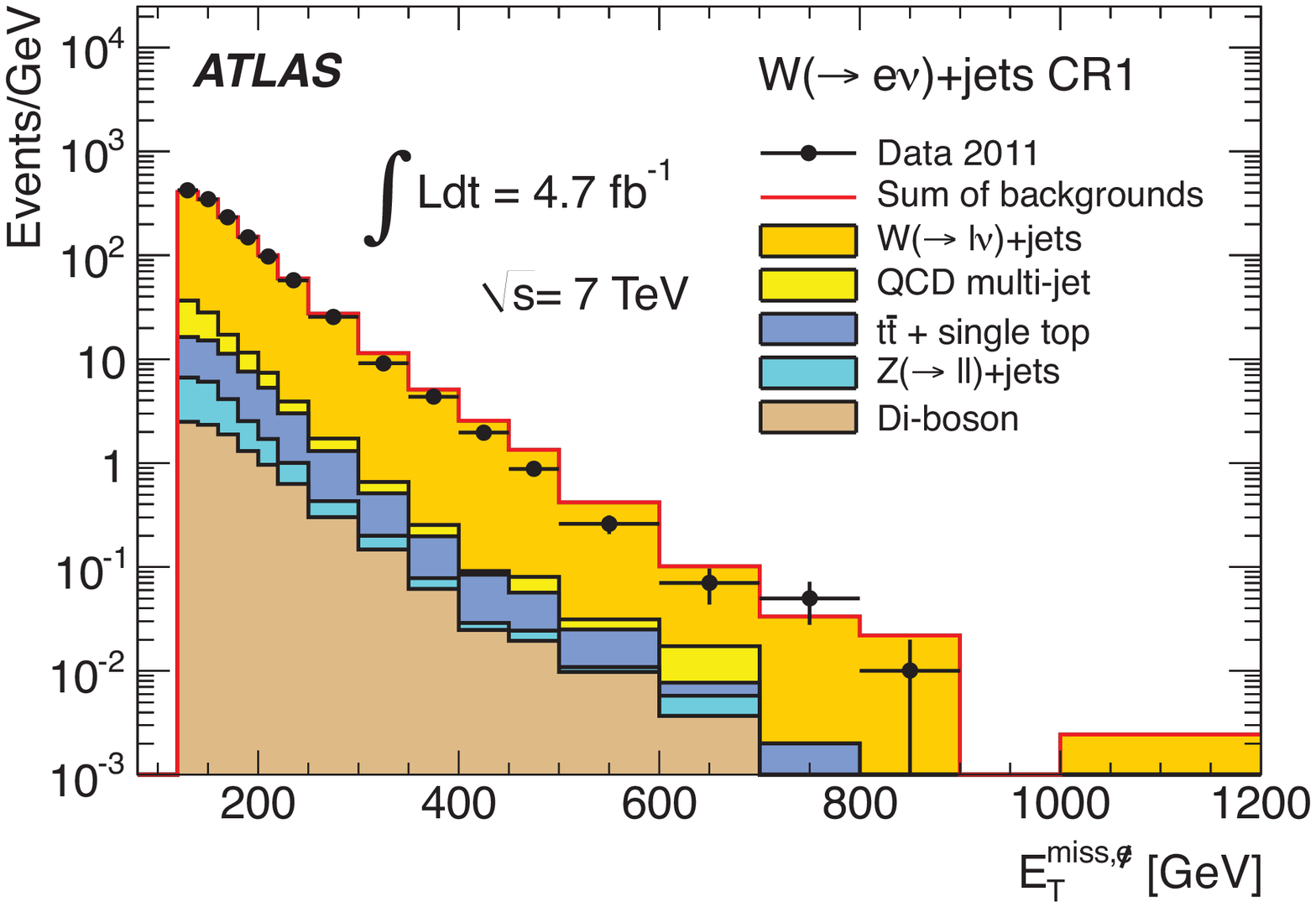}
\includegraphics[width=0.49\linewidth]{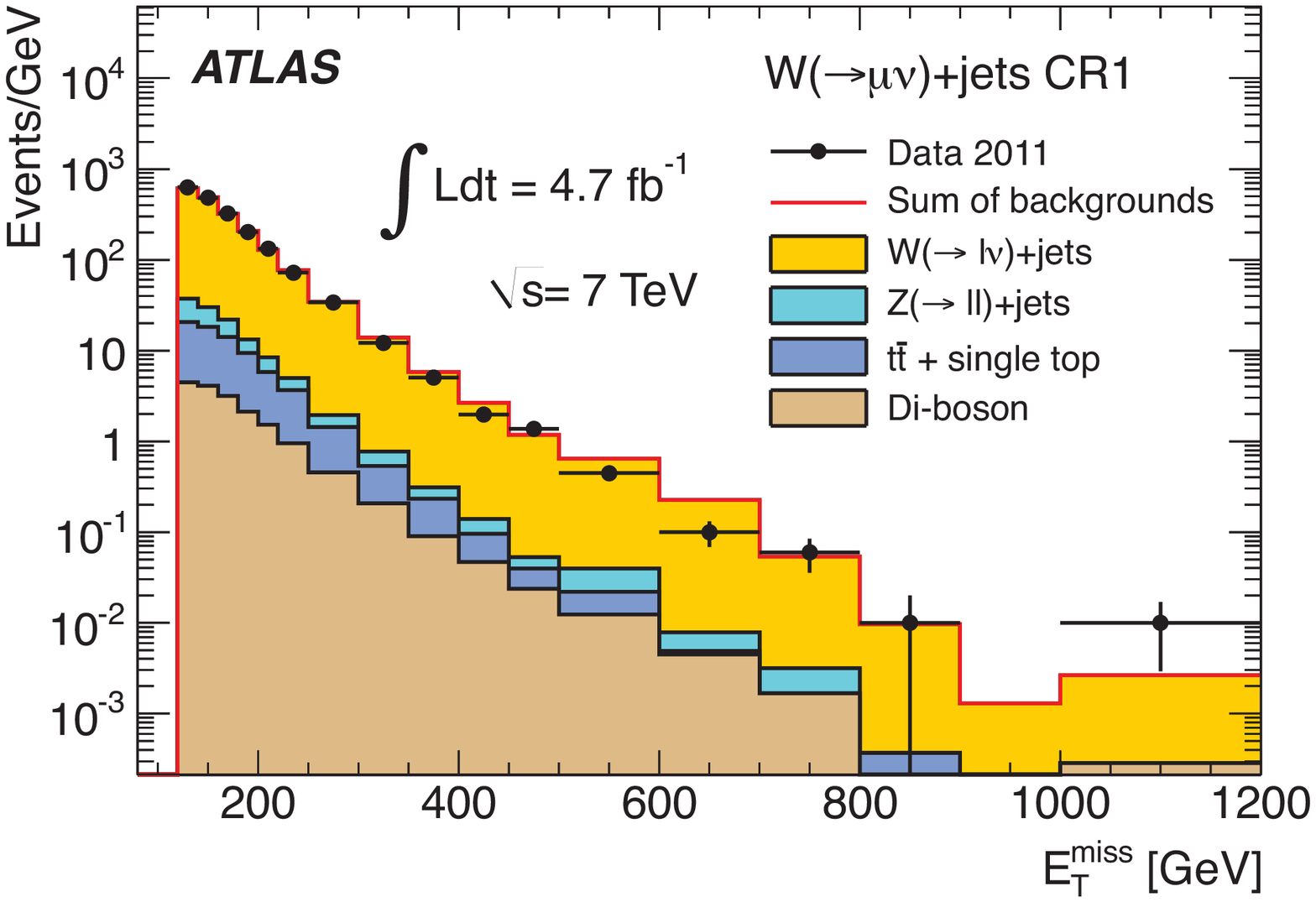}
\caption{Kinematic distributions in the control regions corresponding
  to SR1 (labelled CR1) are shown. The upper row is the leading
  electron and muon \pt\ distribution for $Z\ra \ee$+jets~(left) and
  $Z\ra \mumu$+jets~(right) and shows distributions after SR1 cuts on
  jets and \met.  The lower row is the missing transverse momentum
  distribution $E_{\mathrm{T}}^{\mathrm{miss}, \not e}$ for $W\ra
  e\nu$+jets~(left) and \met\ for $W\ra \mu\nu$+jets~(right) also
  after SR1 jet and \met\ cuts.}
\label{fig:cr}
\end{center}
\end{figure}

\clearpage

\section{Results and interpretation}
\label{sec:results}
\begin{figure}[th]
\begin{center}
\includegraphics[width=0.49\linewidth]{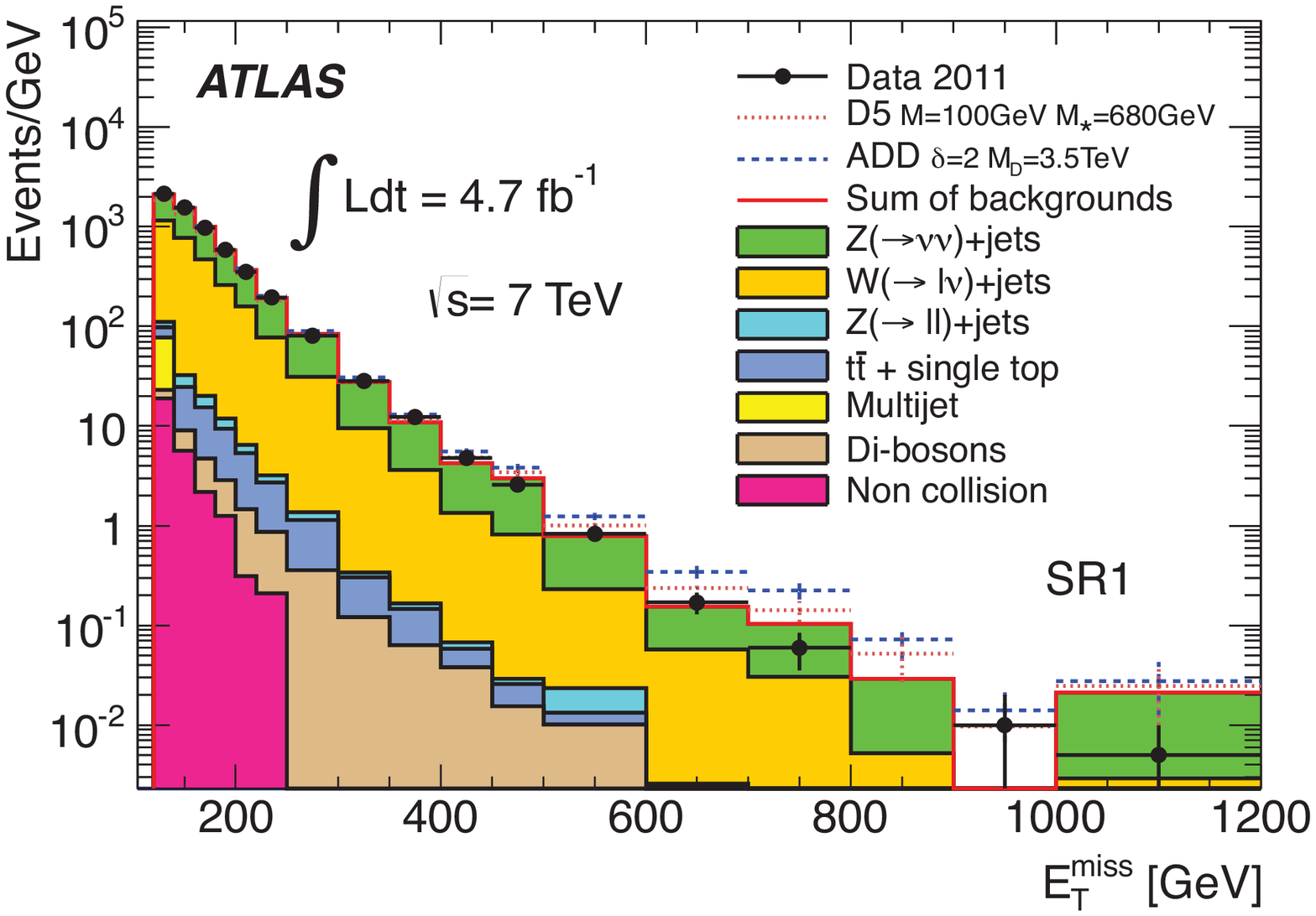}
\includegraphics[width=0.49\linewidth]{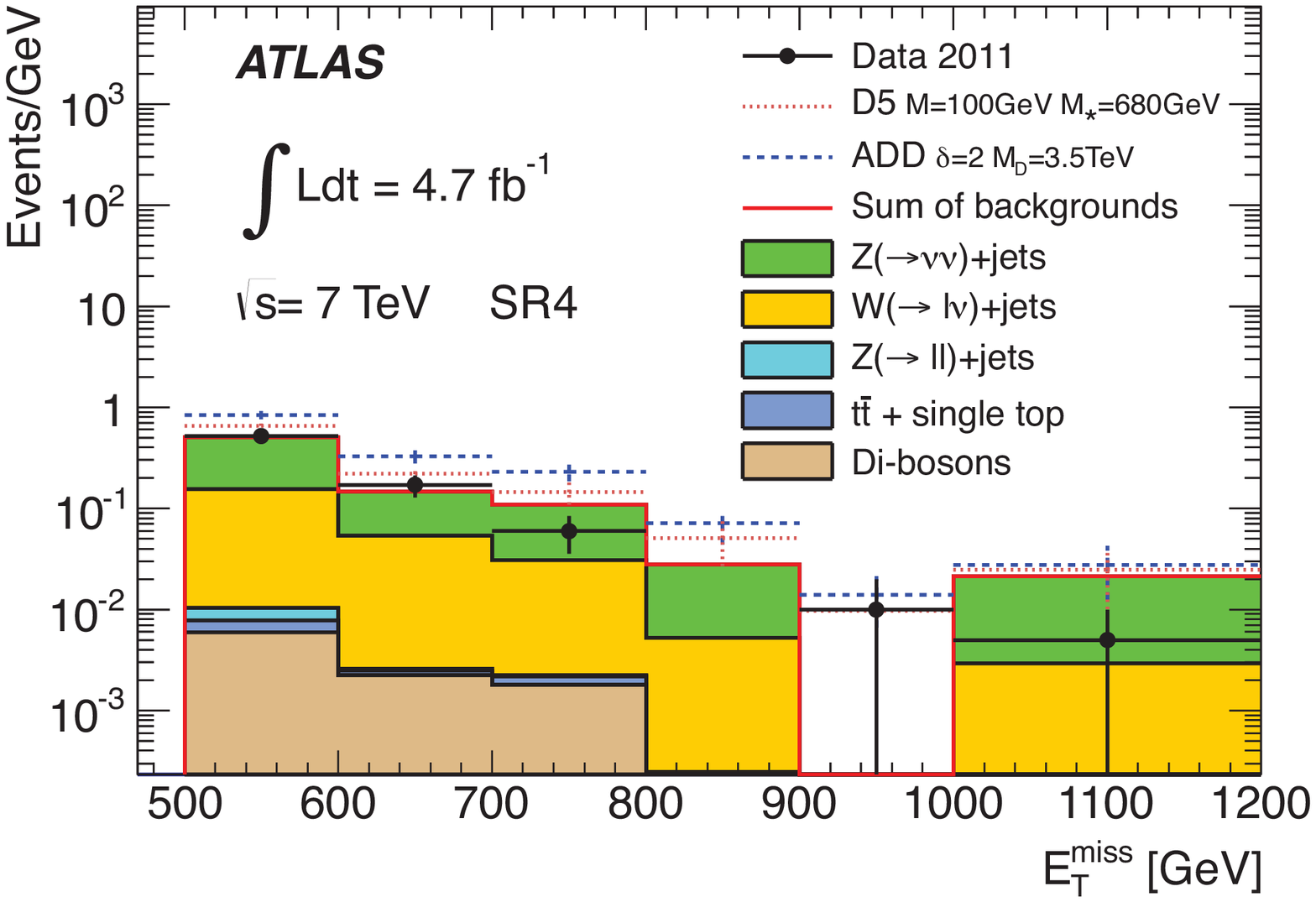}
\includegraphics[width=0.49\linewidth]{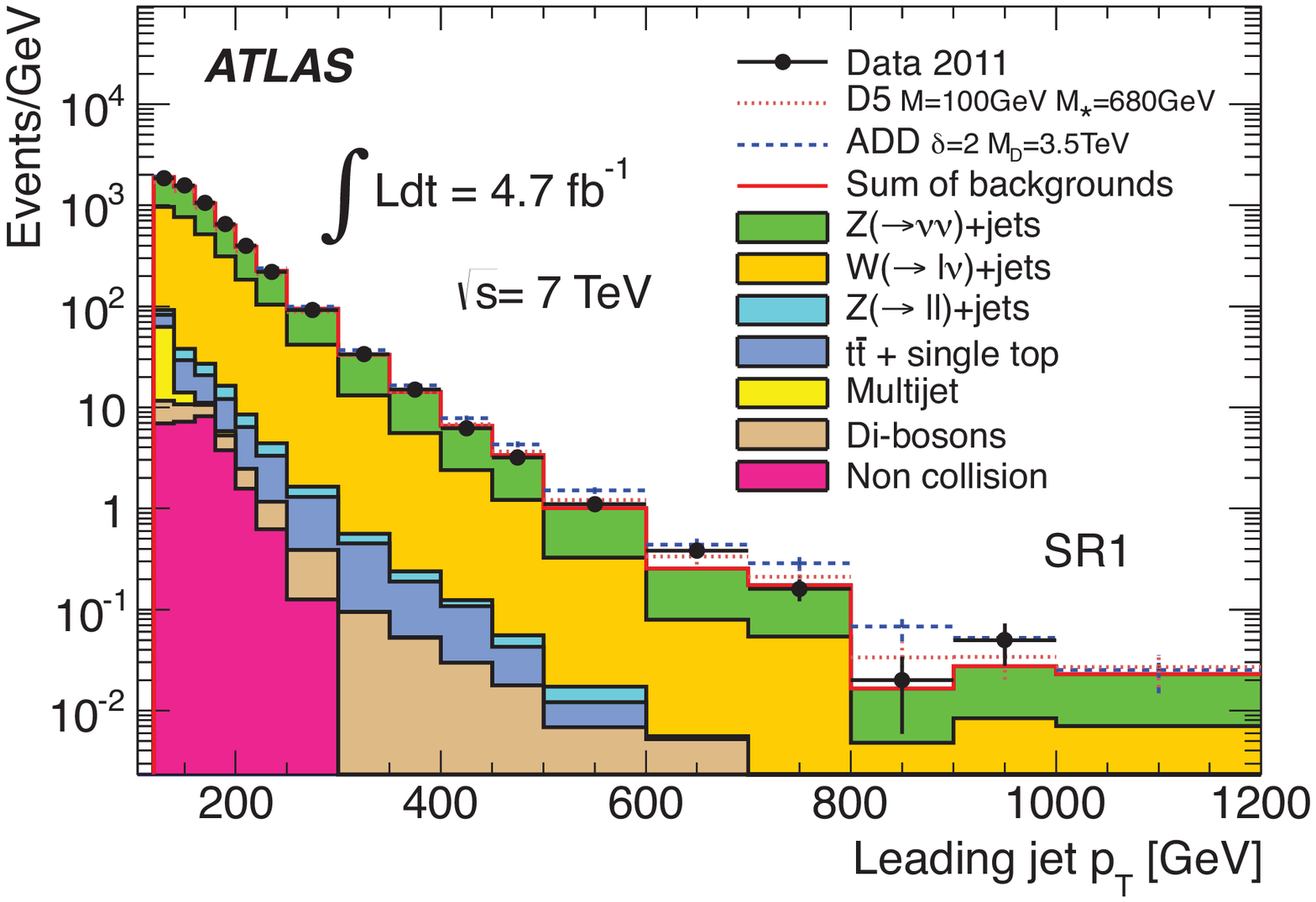}
\includegraphics[width=0.49\linewidth]{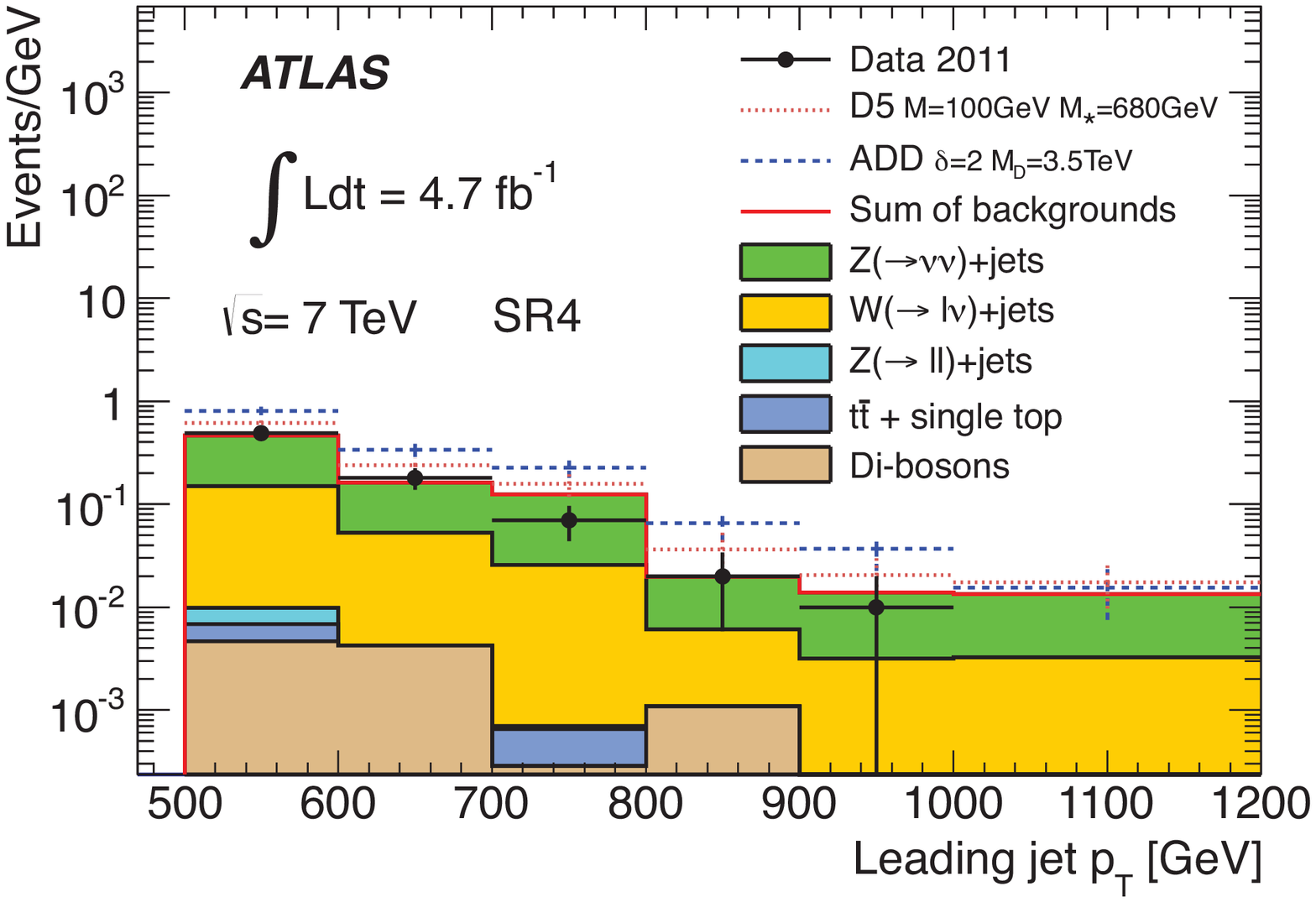}
\includegraphics[width=0.49\linewidth]{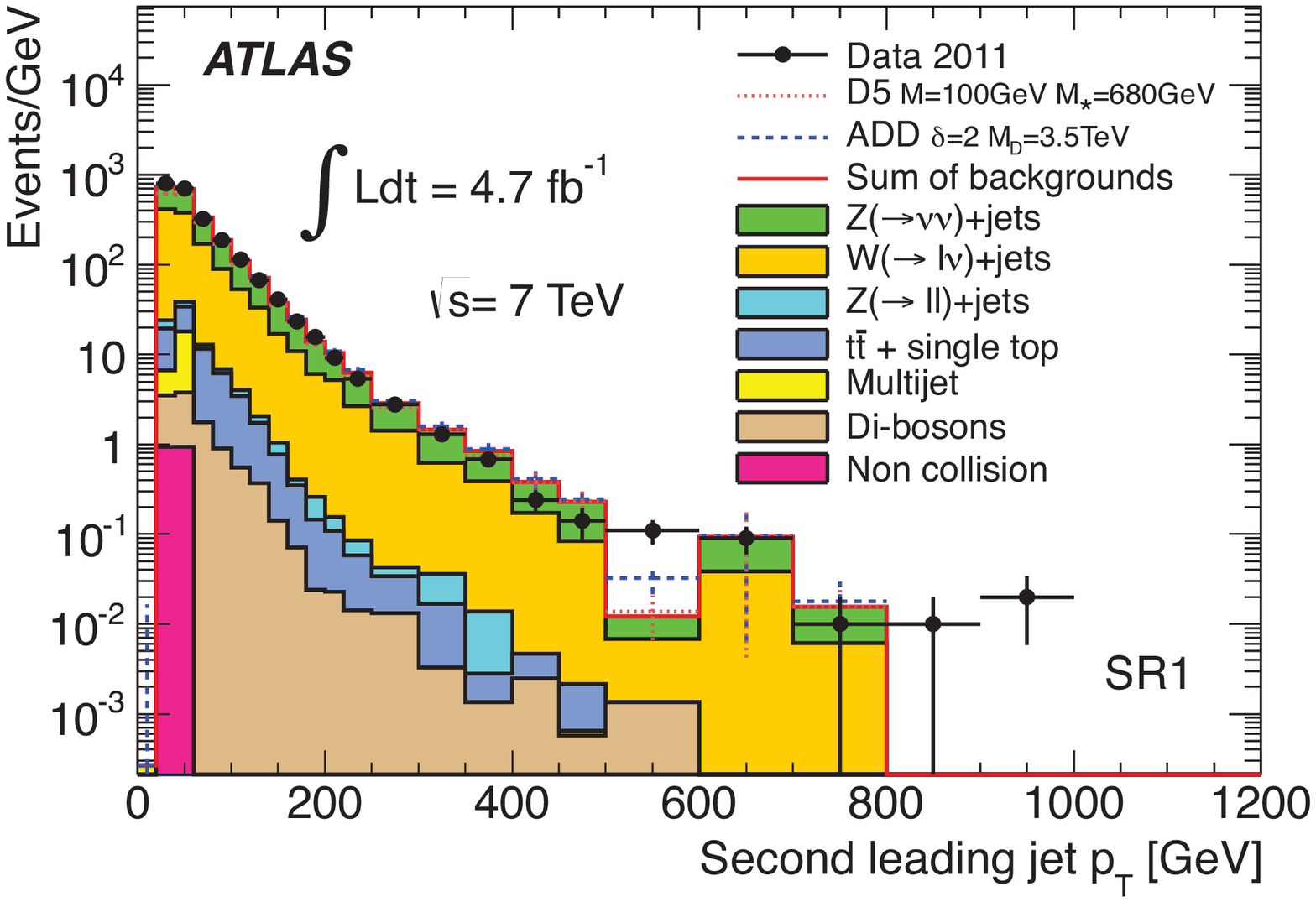}
\includegraphics[width=0.49\linewidth]{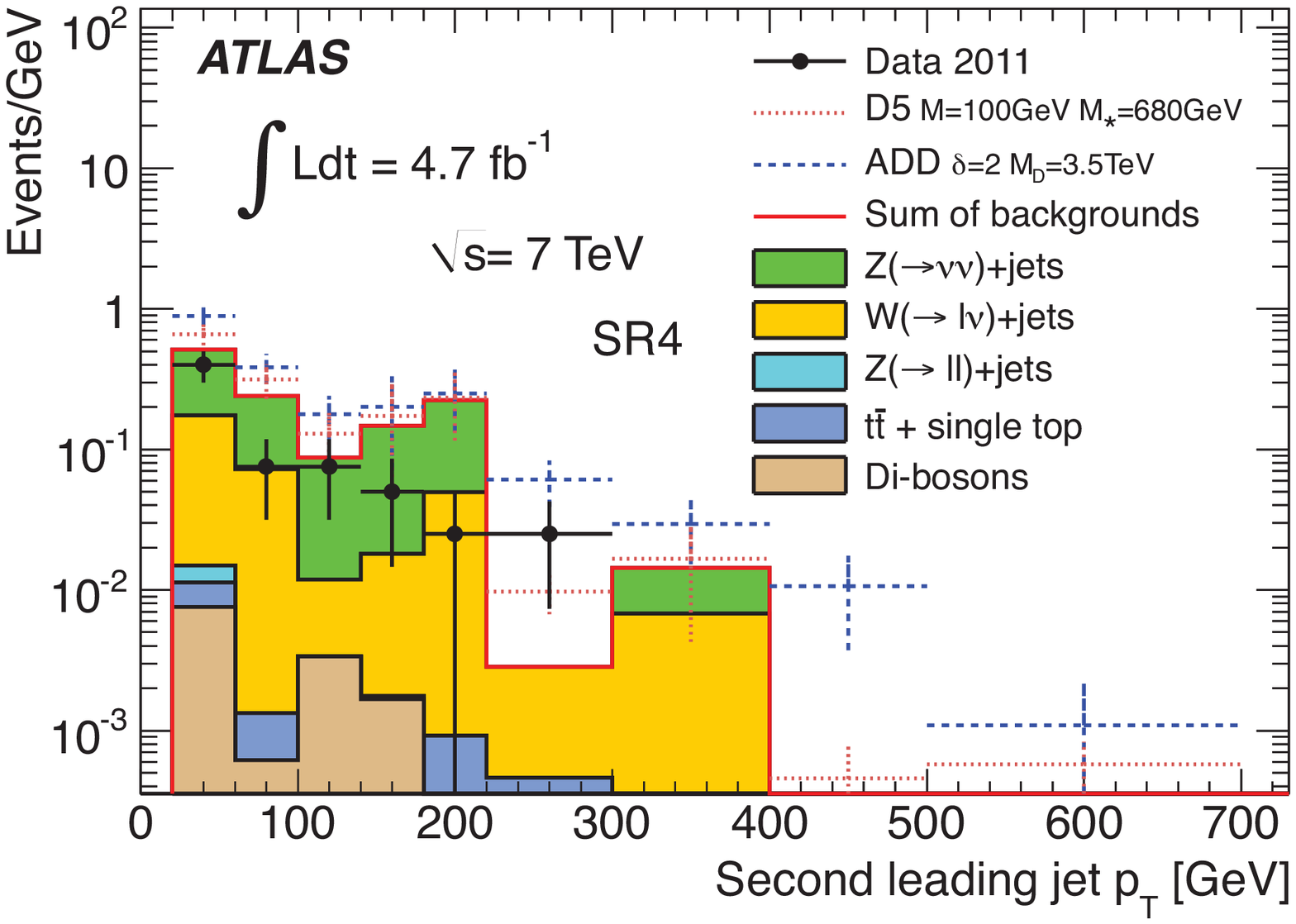}
\caption{Kinematic distributions for signal regions SR1 on the left
  and SR4 on the right. Signal distributions for ADD and WIMP samples
  for cross sections equal to the excluded values are drawn as dashed
  lines on top of the predicted background distributions. The
  electroweak backgrounds (see equation~\ref{eq:bg:2}) are determined
  in bins of the variable that is plotted.}
\label{fig:sr1sr4}
\end{center}
\end{figure}
\begin{table}[tbp]
\begin{center}
\begin{tabular}{|l|cccc|}
  \hline
  Sample & SR1 [ \% ] & SR2 [ \% ] & SR3 [ \% ] & SR4 [ \% ]\\\hline
  $Z\ra \nu\bar{\nu}$+jets& $1.706\pm0.013$ & $0.159\pm0.004$ & $0.0170\pm0.0013$ & $0.0027\pm0.0005$ \\
  ADD, $n=2$ & $30.9\pm0.2$ & $9.2\pm0.1$ & $2.60\pm0.07$ & $0.74\pm0.04$ \\
  ADD, $n=3$ & $33.2\pm0.2$ & $11.7\pm0.1$ & $3.92\pm0.08$ & $1.18\pm0.05$ \\
  ADD, $n=4$ & $34.3\pm0.2$ & $13.8\pm0.1$ & $4.97\pm0.09$ & $1.67\pm0.05$ \\
  ADD, $n=5$ & $35.1\pm0.2$ & $14.5\pm0.1$ & $5.50\pm0.09$ & $2.00\pm0.06$ \\
  ADD, $n=6$ & $35.0\pm0.2$ & $15.0\pm0.2$ & $6.01\pm0.10$ & $2.23\pm0.06$ \\
  D1, $m_\chi=10\GeV$    & $20.5\pm0.3$ & $3.3\pm0.1$ & $0.54\pm0.01$ & $0.09\pm0.01$ \\
  D1, $m_\chi=1000\GeV$  & $32.2\pm0.4$ & $10.3\pm0.2$ & $2.88\pm0.04$ & $0.79\pm0.02$ \\
  D5, $m_\chi=10\GeV$    & $30.4\pm0.4$ & $8.3\pm0.2$ & $2.04\pm0.03$ & $0.52\pm0.01$ \\
  D5, $m_\chi=1000\GeV$  & $36.2\pm0.4$ & $12.6\pm0.2$ & $4.14\pm0.05$ & $1.24\pm0.03$ \\
  D9, $m_\chi=10\GeV$    & $36.9\pm0.5$ & $12.9\pm0.3$ & $4.23\pm0.15$ & $1.31\pm0.08$ \\
  D9, $m_\chi=1000\GeV$  & $37.6\pm0.5$ & $13.9\pm0.3$ & $4.70\pm0.16$ & $1.68\pm0.09$ \\
  D11, $m_\chi=10\GeV$   & $30.3\pm0.4$ & $12.3\pm0.3$ & $4.57\pm0.15$ & $1.52\pm0.09$ \\
  D11, $m_\chi=1000\GeV$ & $33.7\pm0.5$ & $17.0\pm0.3$ & $7.56\pm0.20$ & $3.27\pm0.13$ \\
  \hline
\end{tabular}
\caption{\label{tab:acc} Typical acceptances determined with MC
  simulations for the main background process $Z\ra \nu\bar{\nu}$+jets
  as well as for ADD and selected WIMP samples. For the $Z\ra
  \nu\bar{\nu}$+jets sample at least one parton with a minimum
  transverse momentum of $20\GeV$ is required, for the ADD and WIMP
  samples it is at least one parton with a momentum of $80\GeV$. The
  values are given in percent and errors are statistical only.}
\end{center}
\end{table}
The SM predictions are found to be consistent with the number of
observed events in data for all signal regions considered. Comparisons
of the SM predictions to the measured \met\ and leading and
sub-leading jet \pt\ distributions are shown for SR1 and SR4 in
figure~\ref{fig:sr1sr4}. For illustration, the figures also contain
simulated signal distributions for ADD and WIMP models added to the
total background. Agreement both in the shape and the overall
normalisation between SM predictions and data is observed in all
cases. To facilitate comparisons with other experiments both 90\% and
the more conventional 95\% confidence level (CL) upper limits are
produced. These limits are on the visible cross section defined as
cross section times acceptance and efficiency ($\sigma \times A \times
\epsilon$) and they are based on the modified frequentist
$CL_\mathrm{s}$ prescription~\cite{Read:2002hq}. The limits are
derived by comparing the probabilities, based on Poisson
distributions, that the observed number of events is compatible with
the SM and the SM-plus-signal expectations.  The mean values of the
Poisson distributions are determined by the signal prediction, plus
contributions from background processes extrapolated from the CRs to
each SR.  The number of events is integrated over the whole SR.
Expected limits are obtained by repeating the analysis with
pseudo-data obtained from Monte Carlo simulations. The distributions
of the simulated probabilities for many pseudo-experiments allow $\pm
1 \sigma$ bands to be plotted for the expected values. Systematic
uncertainties (and their correlations) associated with SM backgrounds
and the integrated luminosity are taken into account via nuisance
parameters using a profile likelihood
technique~\cite{Cowan:2010js}. The nuisance parameters are assumed to
be Gaussian distributed in the likelihood fit. The resulting visible
cross-section limits, which apply for any source of BSM events, are
summarised in table~\ref{tab:bg}. Typical efficiencies of selection
criteria related to jets and \met\ of $\epsilon \sim 83\%$ are found
in simulated $Z\ra \nu\bar{\nu}$+jets, WIMP or ADD samples. Typical
acceptances are given in table~\ref{tab:acc}. The negative search
results are interpreted in terms of limits on ADD and WIMP model
parameters in the following sections.

\subsection{Large extra dimensions}
\begin{figure}[htb]
\begin{center}
\includegraphics[width=0.49\linewidth]{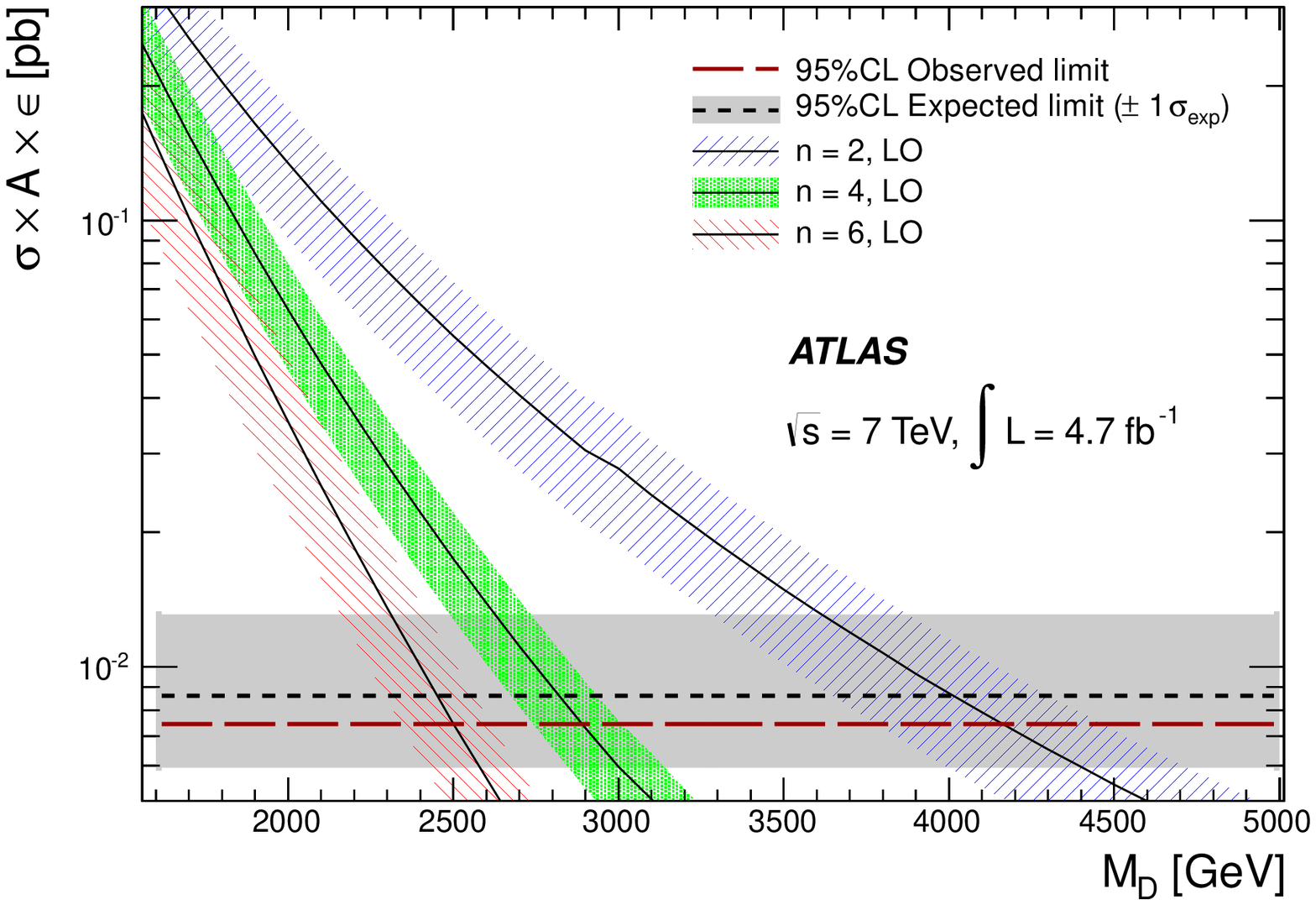}
\includegraphics[width=0.49\linewidth]{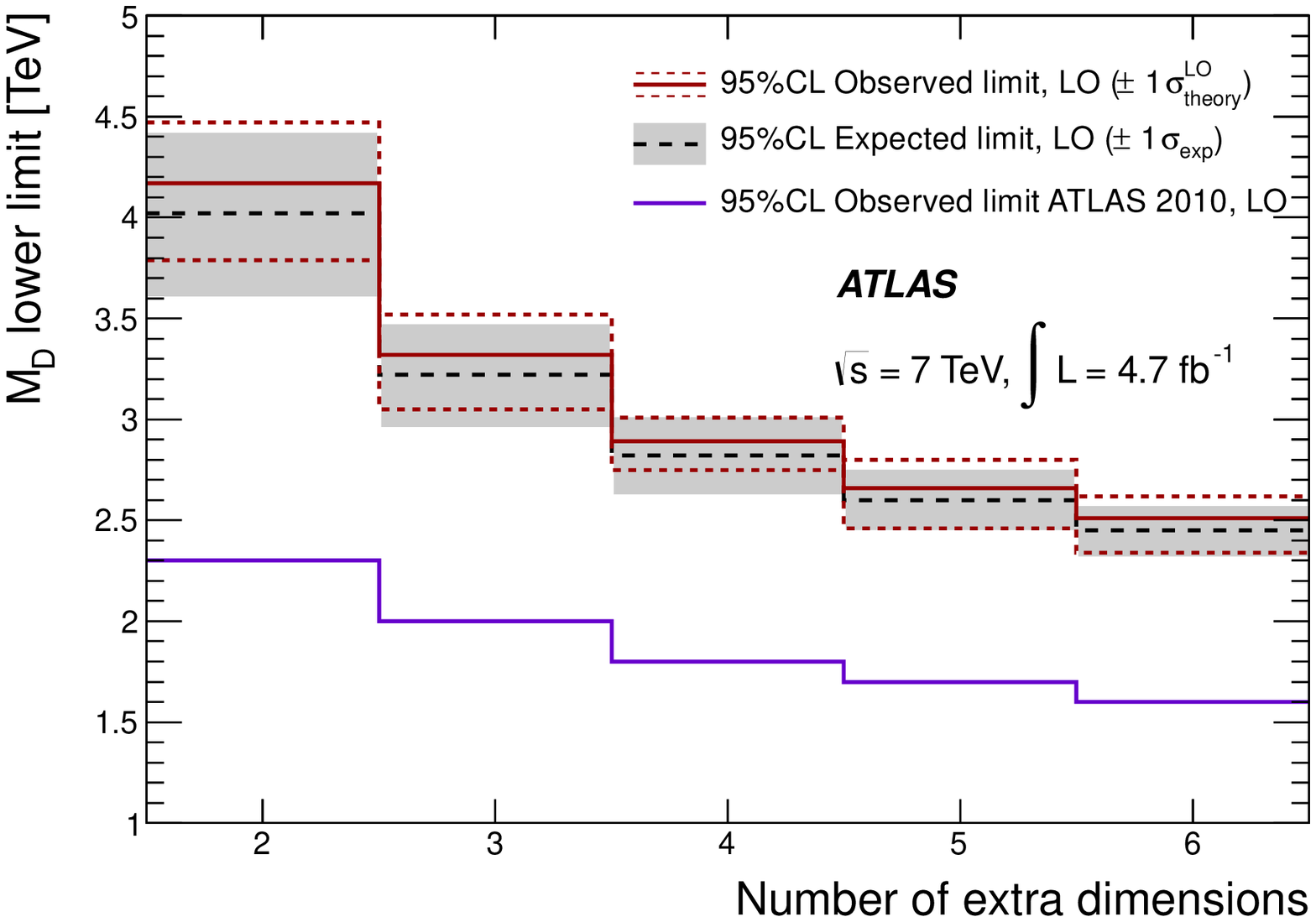}
\caption{{\bf Left:} Visible cross sections in SR4 as a function of
  $M_\mathrm{D}$ as predicted by the effective ADD theory, for
  $n=2,4,6$ extra dimensions. The coloured bands correspond to the
  theoretical systematic uncertainties (PDF, ISR/FSR, scale). The
  horizontal lines are the expected and observed cross-section limits
  at 95\% CL, taking into account experimental systematic
  uncertainties fully correlated between signal and background, as
  well as uncertainties on the luminosity estimate, trigger
  efficiency, and MC statistical uncertainties. The inclusion of
  signal uncertainties here increases the cross-section limits
  compared to those given in table~\ref{tab:bg}, which exclude signal
  uncertainties. {\bf Right:} 95\% CL lower limits on $M_\mathrm{D}$
  for different numbers of extra dimensions based on SR4. Observed and
  expected limits including all but the theoretical signal
  uncertainties are shown as solid and dashed lines, respectively. The
  grey $\pm 1 \sigma$ band around the expected limit is the variation
  expected from statistical fluctuations and experimental systematic
  uncertainties on SM and signal processes. The impact of the
  theoretical uncertainties is shown by the red small-dashed $\pm 1
  \sigma$ limits. The previous ATLAS limit~\cite{Aad:2011xw} is also
  shown for comparison.}
\label{fig:add}
\end{center}
\end{figure}
Experimental and theoretical systematic uncertainties that affect the
ADD signal are considered in order to set limits on the model
parameters. The experimental uncertainties on JES, JER, and \met\ are
considered to be fully correlated with those obtained for the
background estimate. They range from 3--10\% depending on the signal
region and the number of extra dimensions. An additional 1\%
uncertainty on the trigger, and a 3.9\% uncertainty on the luminosity
are also considered for the signal simulation only.  Theoretical
uncertainties on the expected ADD signal are associated with the PDF
set, ISR/FSR, and the factorisation and renormalisation scales.  For
the PDF uncertainties, the {\tt CTEQ6.6} error sets are used,
converted from 90\% to 68\% CL. They range from 4--14\% on the product
of signal cross section and acceptance $(\sigma \times A)$, depending
on the number of extra dimensions. Uncertainties coming from the
modelling of ISR/FSR are determined by varying the simulation
parameters of {\tt PYTHIA} within a range that is consistent with
experimental data~\cite{Skands:2010ak}. The resulting uncertainties
vary from about 3--14\%. The dominant theoretical systematic
uncertainty affecting mostly the cross section rather than the
acceptance is from the factorisation and renormalisation
scales. Varying these scales between twice and half their default
values, following common practice, results in 20--30\% uncertainties
on $\sigma \times A$.\footnote{Note that in ref.~\cite{Aad:2011xw} the
  \emph{squared} factorisation and renormalisation scales were varied
  between twice and half their default values.}

The visible cross sections predicted by the ADD generator for SR4 are
shown for $n=2,4,6$ extra dimensions as a function of $M_\mathrm{D}$
on the left-hand side of figure~\ref{fig:add}. Theoretical systematic
uncertainties are shown as coloured bands around the cross-section
curves. The 95\% CL expected and observed limits on the visible cross
section $\sigma \times A \times \epsilon$ are shown as horizontal
lines. The effect of restricting the simulated phase space to the
kinematic region where the ADD effective field-theory implementation
is valid is probed by evaluating the cross section after discarding
events for which the parton centre-of-mass energy $\hat{s} >
M^2_\mathrm{D}$. The amount by which the truncated cross sections
differ from the full ones provides a measure for the reliability of
the effective field theory. This difference increases from SR1 to SR4
and with the number of extra dimensions. While the model with $n=2$
extra dimensions is found to be insensitive to truncation effects for
$M_\mathrm{D}$ values near the resulting limits for all signal
regions, $n=3$, 4, 5, and 6 extra dimensions show differences of 2\%,
10\%, 30\%, and 50\% between full and truncated cross sections for SR4
and $M_\mathrm{D}$ values close to the actual
limits~(table~\ref{tab:ADD:limits}). This demonstrates that the high
energy and integrated luminosity used in this search allow to probe
kinematic regions where the effective field-theory model is not
entirely valid.
\begin{table}[tbp]
\begin{center}
\begin{tabular}{|c|cc|cc|cc|}
  \hline
\multirow{2}{*}{$n$}& \multicolumn{2}{c|}{$M_\mathrm{D}$ [ TeV ]} &
\multicolumn{2}{c|}{$R$ [ pm ]} & \multicolumn{2}{c|}{Cross section
  truncation}\\
&LO&NLO&LO&NLO&LO&NLO\\\hline 
2 & 4.17  & 4.37 & $2.8\times 10^7$ & $2.5\times 10^7$ & $0.02\%$& $0.01\%$\\
3 & 3.32  & 3.45 & $4.8\times 10^2$ & $4.5\times 10^2$ & $1.9\%$& $1.3\%$\\
4 & 2.89  & 2.97 & $2.0$ & $1.9$ & $11.8\%$ & $9.9\%$\\
5 & 2.66  & 2.71 & $7.1\times 10^{-2}$ & $7.0\times 10^{-2}$ & $29.5\%$ & $27.2\%$\\
6 & 2.51  & 2.53 & $0.8\times 10^{-2}$ & $0.8\times 10^{-2}$ & $49.1\%$ & $47.9\%$\\\hline
\end{tabular}
\caption{\label{tab:ADD:limits} 95\% CL lower (upper) limits on
  $M_\mathrm{D}$ ($R$) for $n$=2--6 extra dimensions, using a dataset
  corresponding to $4.7\ifb$ at $\sqrt{s} = 7\TeV$. These results are
  obtained using the selection criteria of SR4. All values correspond
  to the nominal observed limits excluding theoretical uncertainties
  in figure~\ref{fig:add}. The last two columns show the relative
  difference between the full cross sections and those of the truncated
  phase space ($\hat{s} < M^2_\mathrm{D}$). The ADD cross sections are
  calculated at both LO and NLO, and the limits are derived from the full,
  not the truncated, phase space.}
\end{center}
\end{table}

The 95\% CL lower limits on $M_\mathrm{D}$ versus $n$ for the full
phase space, not the truncated one, are shown for SR4 on the
right-hand side of figure~\ref{fig:add}. The selection criteria of SR4
provide the best expected limits and are therefore used here. Limits
from SR1, SR2, SR3 are typically 35\%, 15\%, 5\% worse,
respectively. The expected and observed limits in figure~\ref{fig:add}
are produced taking all but the theoretical uncertainties into
account. The grey $\pm 1 \sigma$ band around the expected limit shows
the variation anticipated from statistical fluctuations and from
experimental systematic uncertainties on background and signal
processes. The impact of the theoretical uncertainties associated with
PDFs, ISR/FSR, and factorisation and renormalisation scales is
represented in the right-hand panel by dashed $\pm 1 \sigma$ lines on
either side of the observed limit. The resulting limit is taken as the
observed line excluding theoretical uncertainties.\footnote{The
  previous ATLAS monojet search~\cite{Aad:2011xw} has determined ADD
  parameter limits in a slightly different way. The effect of the
  signal cross section theoretical uncertainty was folded into the
  quoted limit and was not shown separately.} All limits from SR4 are
summarised in table~\ref{tab:ADD:limits}, where the lower (upper)
limits on $M_\mathrm{D}$ ($R$) are shown for cross sections calculated
at LO and NLO. The $K$-factors (defined as
$\sigma_\mathrm{NLO}/\sigma_\mathrm{LO}$) for $n=2$, 3, 4, 5, 6 extra
dimensions are 1.20, 1.20, 1.17, 1.13, 1.09, respectively, and have
been derived for the selection criteria of SR4 by the authors of
ref.~\cite{Karg:2009xk}. $M_\mathrm{D}$ values below $4.17~(4.37)\TeV$
for $n=2$ and $2.51~(2.53)\TeV$ for $n=6$ are excluded at 95\% CL at
LO (NLO).

\subsection{WIMP pair production}
Systematic uncertainties on WIMP pair production are treated similarly
to those of the ADD limits, except for the PDF and ISR/FSR
uncertainties. The former are determined using {\tt CTEQ6M} error sets
for the relative uncertainty around the {\tt CTEQ6L1} central
value. The ISR/FSR uncertainties are estimated differently in a way
that is appropriate for the high-\pt\ ISR/FSR regime probed here: a
WIMP pair recoils against a high-\pt\ ISR/FSR jet, whereas for ADD,
additional low-\pt\ ISR/FSR jets dominate the uncertainty due to the
impact of the jet veto.

The JES/JER/\met\ experimental uncertainties lead to 1--20\%
uncertainties on the WIMP event yield depending on the signal region
and the effective operator considered. Other experimental
uncertainties affecting the WIMP event yield are associated with the
trigger efficiency (1\%) and the luminosity measurement (3.9\%). The
ISR/FSR uncertainties are estimated by varying the jet matching scale
between {\tt MADGRAPH5} and {\tt PYTHIA} by a factor of one half and
two. Moreover, the $\alpha_s$ scale is varied in {\tt PYTHIA} within a
range that is consistent with experimental
data~\cite{Skands:2010ak}. The resulting uncertainties on $\sigma
\times A$, added in quadrature, range from 3--5\% for the matching
scale and 4--6\% for $\alpha_s$ depending on the signal region. A
negligible dependence of the ISR/FSR uncertainties on the choice of
effective operator is found. PDF uncertainties impact mostly the
signal cross section and hardly the acceptance. They are found to
depend on the effective operator chosen and not the particular signal
region (since overall cross-section differences affect the signal
regions in the same way). Uncertainties ranging from 4\% and 5\% for
operators D9 and D5 to 16\% and 18\% for D11 and D1 are found. As for
the ADD model, the dominating theoretical systematic uncertainty is
from the factorisation and renormalisation scales. Varying these
scales between twice and half their default value results in 30\%
signal uncertainties, independent of the effective operator choice or
the signal region.

\begin{figure}[tbp]
\begin{center}
\includegraphics[width=0.49\linewidth]{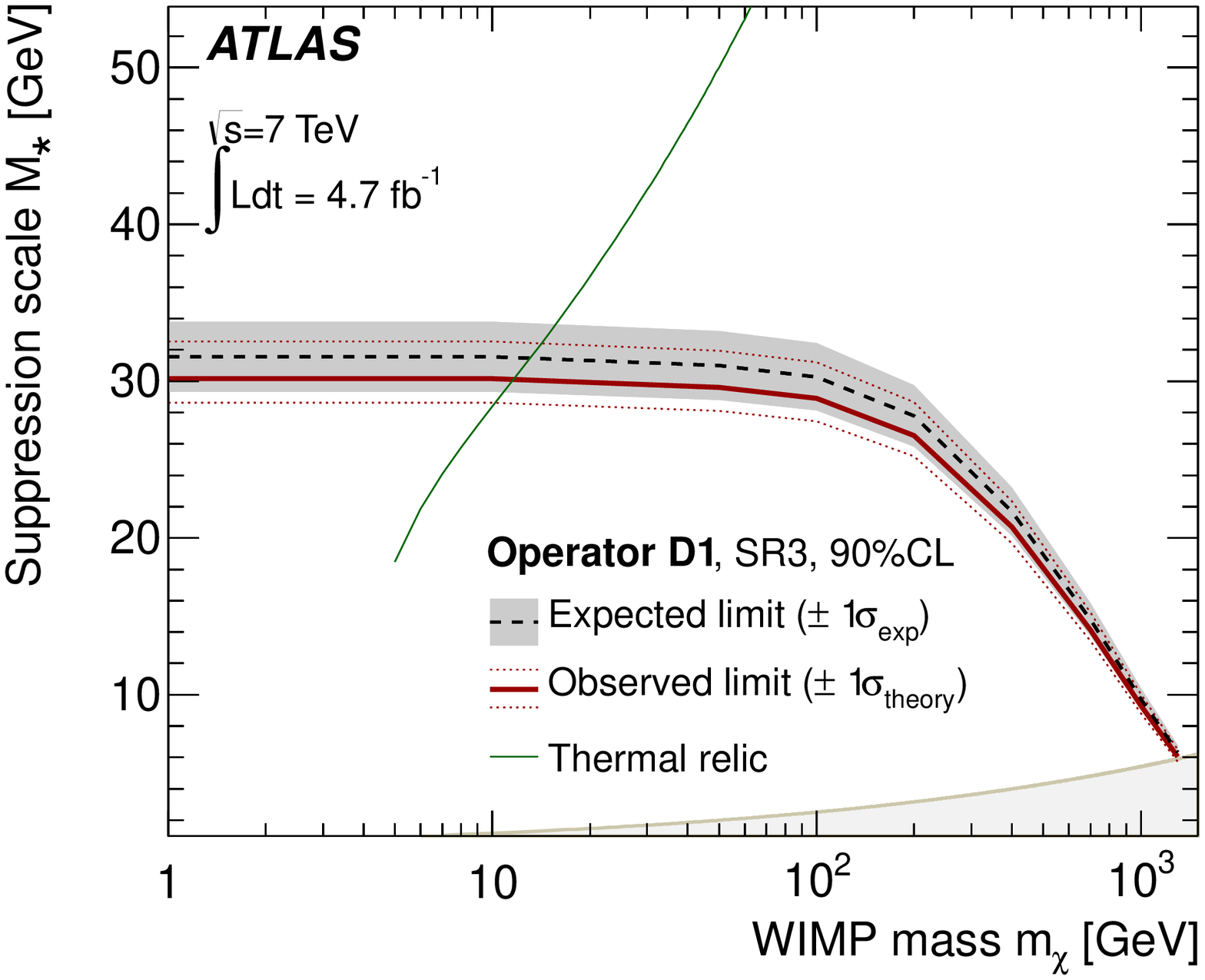}
\includegraphics[width=0.49\linewidth]{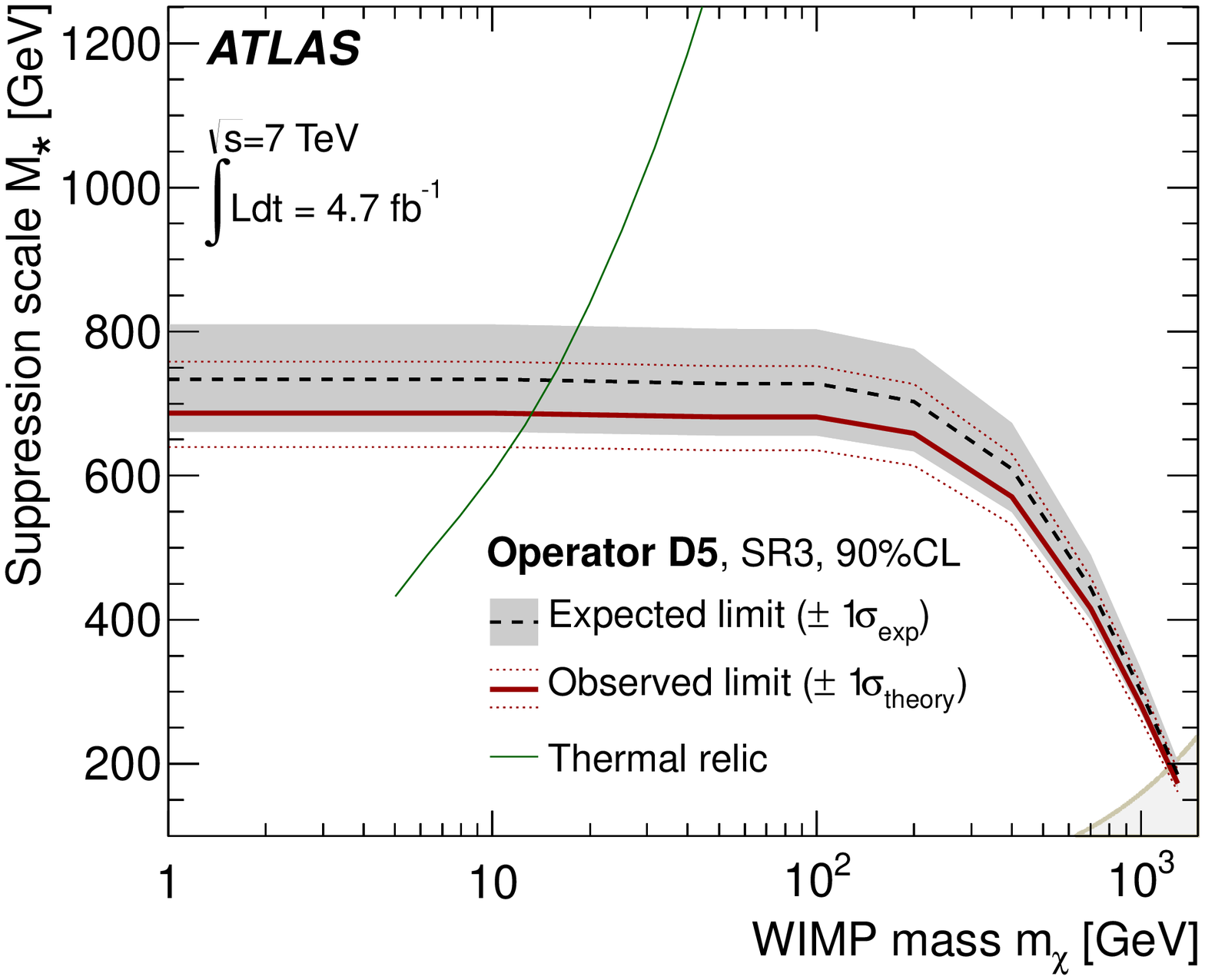}
\includegraphics[width=0.49\linewidth]{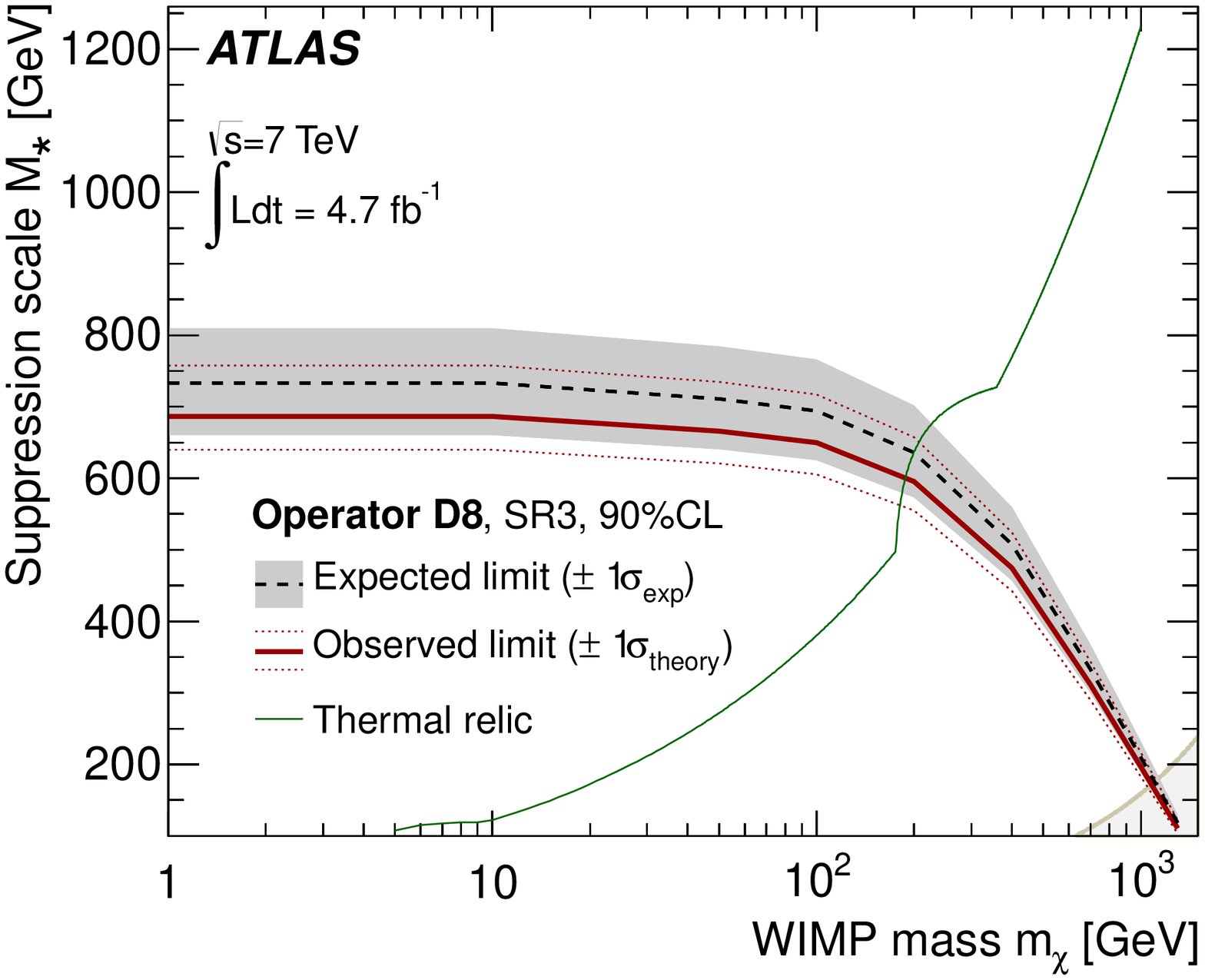}
\includegraphics[width=0.49\linewidth]{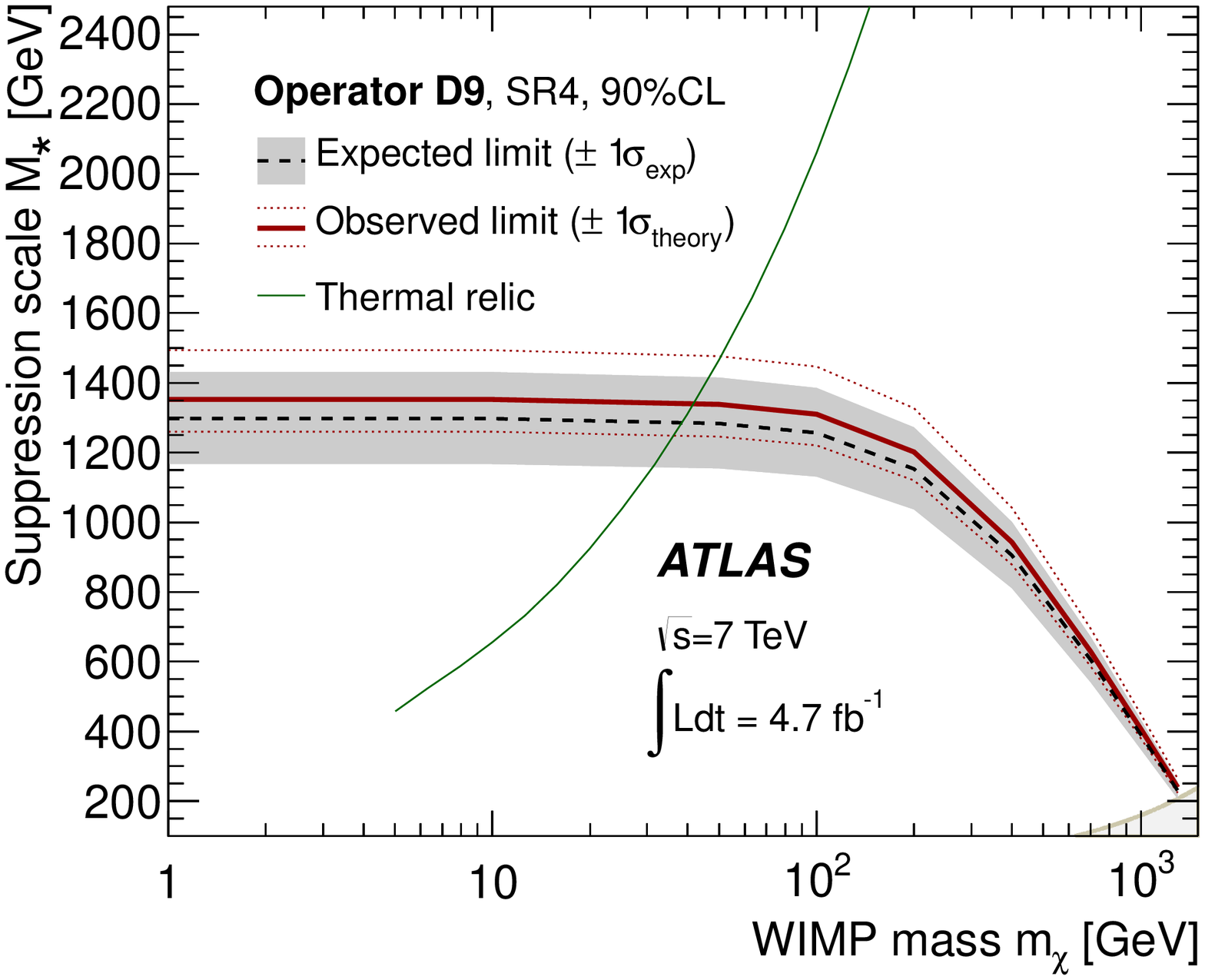}
\includegraphics[width=0.49\linewidth]{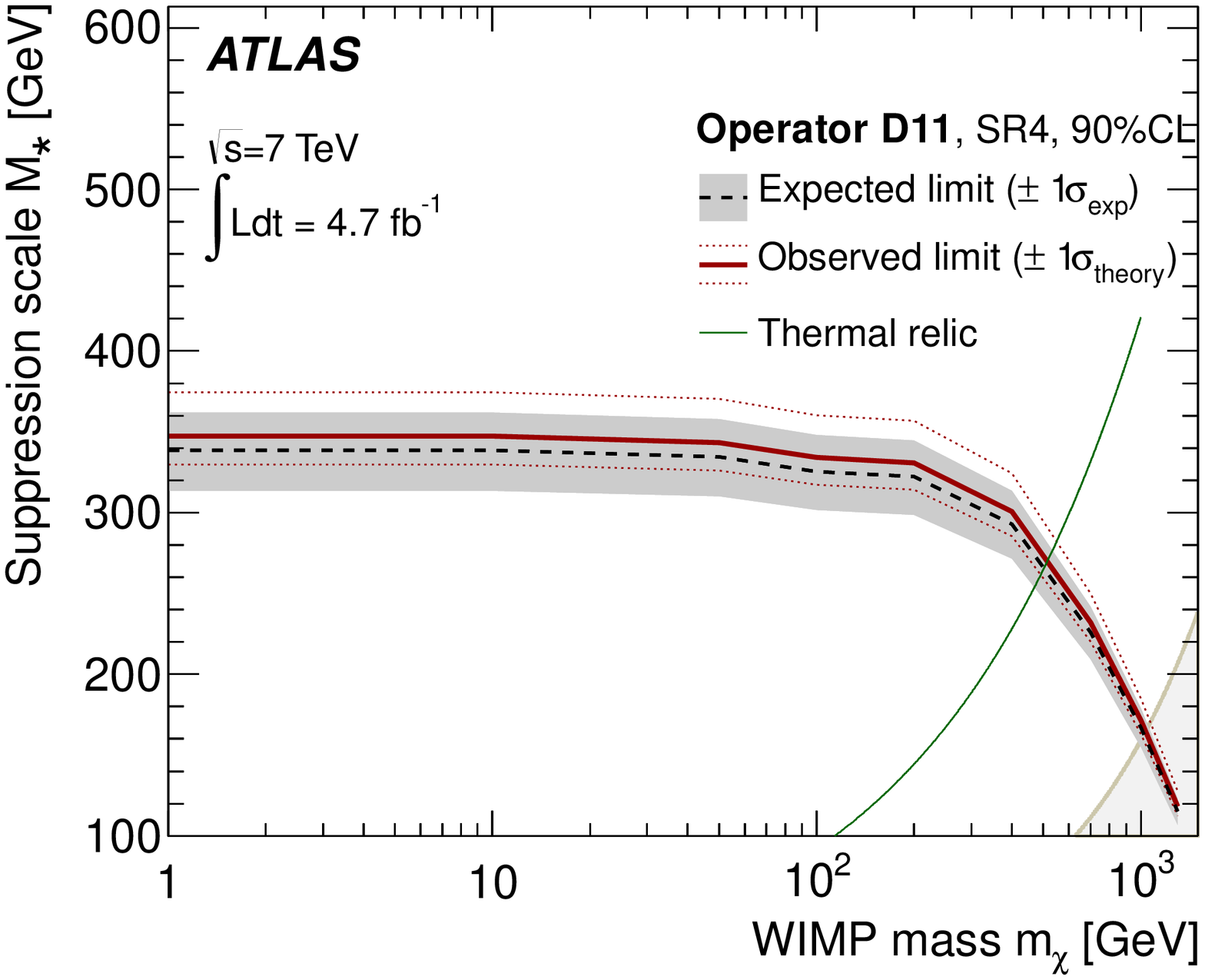}
\caption{ATLAS lower limits at 90\% CL on $M_*$ for different masses
  of $\chi$---the region below the limit lines is excluded. The 90\%
  instead of the 95\% CL lower limits are plotted because the former
  are used in the following figures~\ref{fig:wimp:2} and
  \ref{fig:wimp:3}. Observed and expected limits including all but the
  theoretical signal uncertainties are shown as dashed black and red
  solid lines, respectively. The grey $\pm 1 \sigma$ band around the
  expected limit is the variation expected from statistical
  fluctuations and experimental systematic uncertainties on SM and
  signal processes. The impact of the theoretical uncertainties is
  shown by the thin red dotted $\pm 1 \sigma$ limit lines around the
  observed limit. The $M_*$ values at which WIMPs of a given mass
  would result in the required relic abundance are shown as rising
  green lines~(taken from \cite{Goodman:2010ku}), assuming
  annihilation in the early universe proceeded exclusively via the
  given operator.  The shaded light-grey regions in the bottom right
  corners indicate where the effective field theory approach breaks
  down~\cite{Goodman:2010ku}. The plots for D1, D5, D8 are based on
  SR3, those for D9 and D11 on SR4.}
\label{fig:wimp:1}
\end{center}
\end{figure}
 Figure~\ref{fig:wimp:1} shows the 90\% CL lower limits on the
suppression scale $M_*$, for all operators probed as a function of
WIMP mass $m_\chi$. These limits on $M_*$ are derived from the
cross-section limits at a given mass $m_\chi$. The values displayed
are for the signal regions with the best expected limits, where those
limits from SR3 and SR4 are typically within a few percent of each
other, and those from SR2 (SR1) are 15--20\% (40--50\%) smaller than
in SR3 or SR4. The lower limits are based on simulation samples
produced for $m_\chi$ between 10 and $1300\GeV$. Extrapolations are
shown down to $m_\chi=1\GeV$. These are valid (and could be continued
as constants to even smaller $m_\chi$ values entering the warm or hot
dark-matter regime) since there is negligible change in cross section
or kinematic distributions at the LHC for low-mass WIMPs. As before,
the central values of observed and expected limits on $M_*$ are
displayed taking into account experimental but not theoretical
uncertainties. The effect of $\pm 1 \sigma$ variations on the expected
limit due to statistical fluctuations and experimental uncertainties
is shown as a grey band. The impact of the theoretical uncertainties
is represented by dotted red $\pm 1 \sigma$ lines on either side of
the observed limit. The nominal observed limit line excluding
theoretical uncertainties is the final result. All values of the lower
limits on the suppression scale $M_*$ at 90\% and 95\% CL are listed
in table~\ref{tab:WIMP:limits}. For all operators, the lower limits
are flat up to $m_\chi=100\GeV$ and worsen around
$m_\chi=200\GeV$. Note that the $M_*$ limits for D1 are much smaller
due to the inclusion of a factor $m_q/M_*$ in the definition of the
operator~(see table~\ref{table:wimp:operators}).

\clearpage
\begin{table}[tbp]
\begin{center}
\begin{tabular}{|r|ccccc|}\hline
  \multirow{2}{*}{$m_\chi\,$}
  & \multirow{2}{*}{D1} 
  & \multirow{2}{*}{D5}
  & \multirow{2}{*}{D8}
  & \multirow{2}{*}{D9}
  & \multirow{2}{*}{D11} \\
  &&&&&\\\hline
  1 & 30 ( 29 )& 687 ( 658 )&687 ( 658 ) & 1353 ( 1284 )&  347 ( 335 )\\
  5 & 30 ( 29 )& 687  ( 658 )&687 ( 658 ) & 1353 ( 1284 )& 347 ( 335 )\\
  10 & 30 ( 29 )& 687 ( 658 )&687 ( 658 ) & 1353 ( 1284 )&347 ( 335 )\\
  50 & 30 ( 29 )& 682 ( 653 )&666 ( 638 ) & 1338 ( 1269 )&343 ( 331 )\\
  100 & 29 ( 28 )& 681 ( 653 )&650 ( 623 )& 1310 ( 1243 )&334 ( 322 )\\
  200 & 27 ( 26 )& 658 ( 631 )&595 ( 570 )& 1202 ( 1140 )&331 ( 319 )\\
  400 & 21 ( 20 )& 571 ( 547 )&475 ( 455 )& 943 ( 893 )&301 ( 290 ) \\
  700 & 14 ( 14 )& 416 ( 398 )&311 ( 298 )& 629 ( 596 )&232 ( 223 ) \\
  1000 & 9 ( 9 ) & 281 ( 269 )&196 ( 188 )& 406 ( 384 ) &171 ( 165 ) \\
  1300 & 6 ( 6 ) & 173 ( 165 )&110 ( 106 )& 240 ( 227 ) &118 ( 114 ) \\\hline
\end{tabular}
\caption{\label{tab:WIMP:limits} ATLAS 90\% (95\%) CL observed lower
  limits on the suppression scale $M_*$ as a function of WIMP mass
  $m_\chi$. All values are given in \GeV\ and correspond to the
  nominal observed limit excluding theoretical uncertainties. The
  signal regions with the best expected limits are quoted in all
  cases, SR3 is used for D1, D5 and D8, SR4 for D9 and D11.}
\end{center}
\end{table}

The light-grey shaded regions in figure~\ref{fig:wimp:1} indicate
where the effective field theory approach for WIMP pair production
breaks down~\cite{Goodman:2010ku}~(bottom-right corner in all
plots).\footnote{Compared to ref.~\cite{Goodman:2010ku} the valid
  region of D1 shown here accounts for the factor of $m_q$ in the
  definition of D1 (see table~\ref{table:wimp:operators}).}  Except
for some of the $m_\chi = 1300\GeV$ points, the $M_*$ limits set in
this analysis are well above these bounds. No further measures are
taken to ensure that the energy transfer in monojet events in this
dataset remains in the valid region of the effective field
theory. Such a region of validity cannot be defined without precise
knowledge of the BSM physics, over which the effective operators
integrate.

Figure~\ref{fig:wimp:1} also includes \emph{thermal relic}
lines~(taken from \cite{Goodman:2010ku}) which correspond to a
coupling, set by $M_*$, of WIMPs to quarks or gluons such that WIMPs
have the correct relic abundance as measured by the WMAP
satellite~\cite{Komatsu:2010fb}, in the absence of any other
interaction than the one considered. Under the assumption that DM is
entirely composed of thermal relics, ATLAS limits on $M_*$ that are
above the value required for the thermal relic density exclude the
case where DM annihilates exclusively to SM particles via the
corresponding operator. Should thermal relic WIMPs exist in these
regions (above the thermal relic line), there would have to be other
annihilation channels or annihilation via other operators in order to
be consistent with the WMAP measurements.

\begin{figure}[tbp]
\begin{center}
\includegraphics[width=0.8\linewidth]{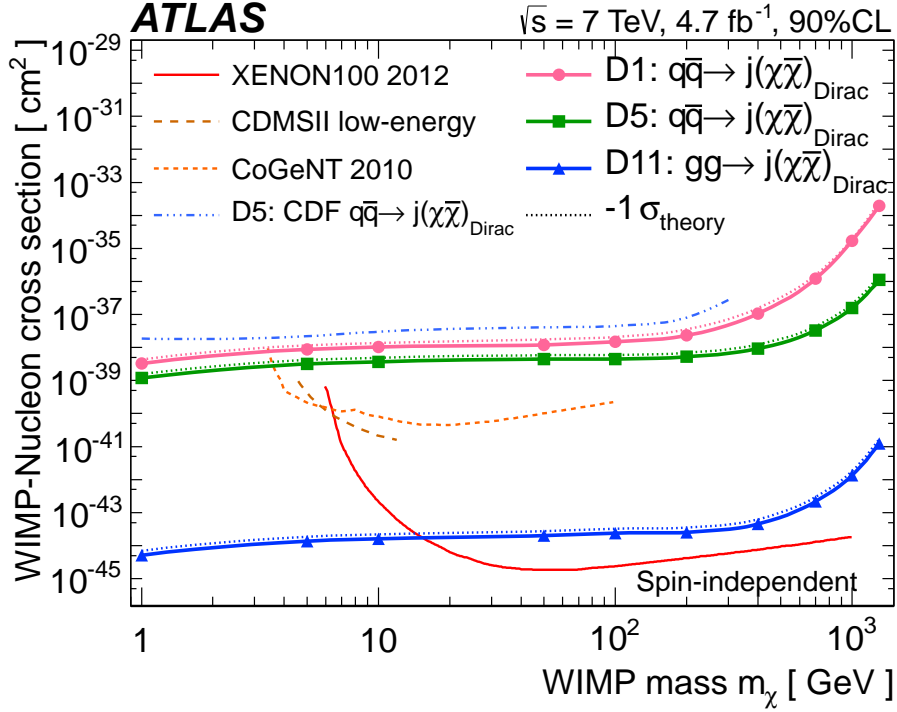}
\caption{Inferred 90\% CL ATLAS limits on spin-independent
  WIMP-nucleon scattering. Cross sections are shown versus WIMP mass
  $m_\chi$. In all cases the thick solid lines are the observed limits
  excluding theoretical uncertainties; the observed limits
  corresponding to the WIMP-parton cross section obtained from the
  $-1\sigma_{\mathrm{theory}}$ lines in figure~\ref{fig:wimp:1} are
  shown as thin dotted lines. The latter limits are conservative
  because they also include theoretical uncertainties. The ATLAS limits
  for operators involving quarks are for the four light flavours
  assuming equal coupling strengths for all quark flavours to the
  WIMPs. For comparison, 90\% CL limits from the
  XENON100~\cite{2012arXiv1207.5988X},
  CDMSII~\cite{PhysRevLett.106.131302}, CoGeNT~\cite{Aalseth:2010vx},
  CDF~\cite{2012PhRvL108u1804A}, and CMS~\cite{CmsPreprint}
  experiments are shown.}
\label{fig:wimp:2}
\end{center}
\end{figure}
 
\begin{figure}[tbp]
\begin{center}
\includegraphics[width=0.8\linewidth]{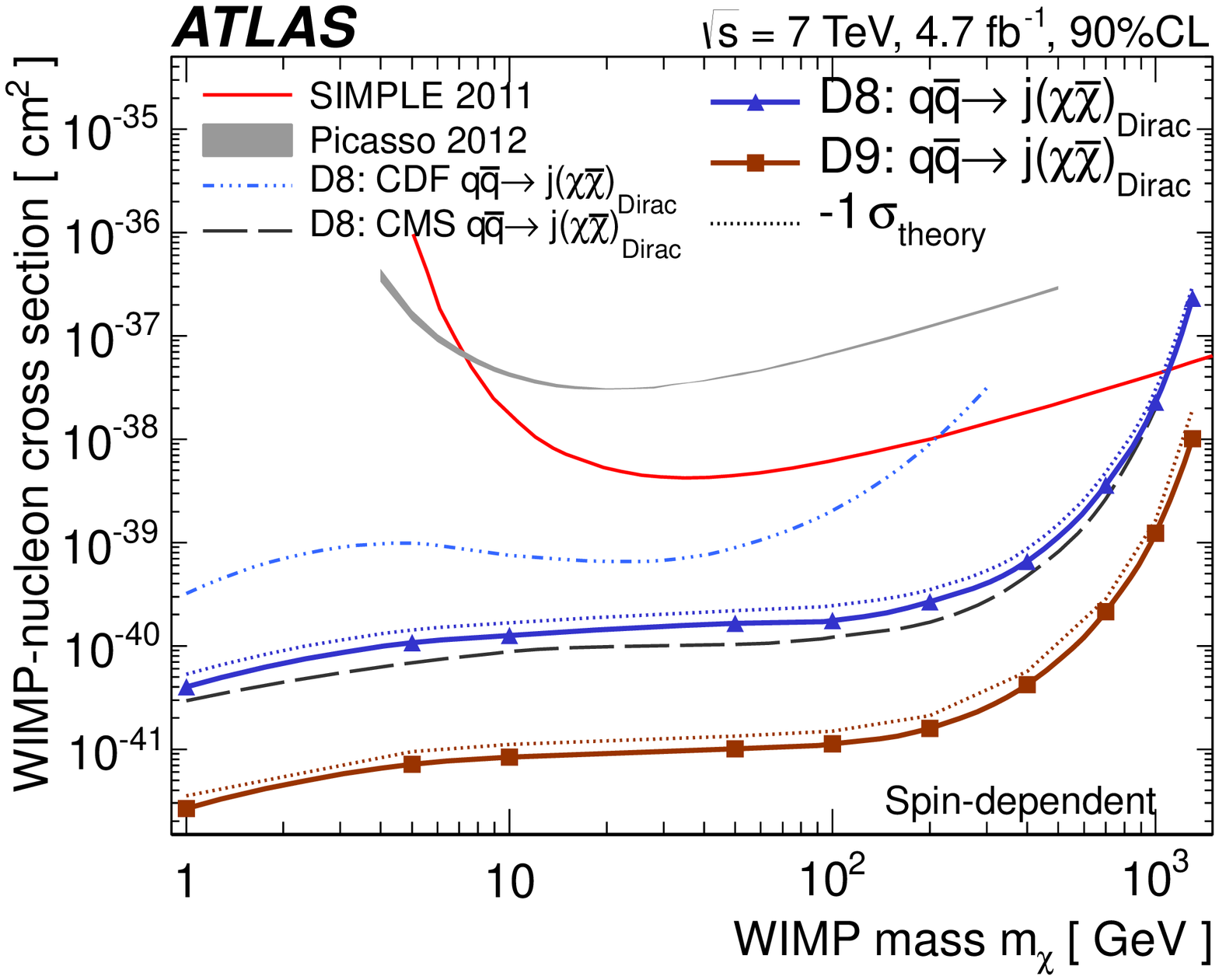}
\caption{Inferred 90\% CL ATLAS limits on spin-dependent WIMP-nucleon
  scattering. Cross sections are shown versus WIMP mass $m_\chi$. In
  all cases the thick solid lines are the observed limits excluding
  theoretical uncertainties, the observed limits corresponding to the
  WIMP-parton cross section obtained from the
  $-1\sigma_{\mathrm{theory}}$ lines in figure~\ref{fig:wimp:1} are
  shown as thin dotted lines.  The latter limits are conservative
  because they also include theoretical uncertainties. The ATLAS
  limits are for the four light flavours assuming equal coupling
  strengths for all quark flavours to the WIMPs. For comparison, 90\%
  CL limits from the SIMPLE~\cite{Felizardo:2011uw},
  Picasso~\cite{Archambault:2012pm}, CDF~\cite{2012PhRvL108u1804A},
  and CMS~\cite{CmsPreprint} experiments are shown.}
\label{fig:wimp:3}
\end{center}
\end{figure}

In the effective operator approach, the ATLAS bounds on $M_*$ for a
given $m_\chi$ can be converted to bounds on WIMP-nucleon scattering
cross sections, which are probed by direct dark matter detection
experiments. These bounds describe scattering of WIMPs from nucleons
at a very low momentum transfer of the order of a keV. Depending on
the type of interaction, contributions to \emph{spin-dependent} or
\emph{spin-independent} WIMP-nucleon interactions are expected. The
translation of ATLAS limits to bounds on WIMP-nucleon scattering cross
sections is done using equations~(3)~to~(6) of
ref.~\cite{Goodman:2010ku}, and the results are shown in
figures~\ref{fig:wimp:2} and ~\ref{fig:wimp:3}.\footnote{There is a
  typographical error in equation~(5) of ref.~\cite{Goodman:2010ku}
  (cross sections for D8 and D9). Instead of $9.18 \times
  10^{-40}\mathrm{cm}^2$ the pre-factor should be $4.7 \times
  10^{-39}\mathrm{cm}^2$.}  As in ref.~\cite{Goodman:2010ku}
uncertainties on hadronic matrix elements are neglected here. The
spin-independent ATLAS limits in figure~\ref{fig:wimp:2} are
particularly relevant in the low $m_\chi$ region ($< 10\GeV$) where
the XENON100~\cite{2012arXiv1207.5988X},
CDMSII~\cite{PhysRevLett.106.131302} or CoGeNT~\cite{Aalseth:2010vx}
limits suffer from a kinematic suppression. Should DM particles couple
exclusively to gluons via D11, the collider limits would be
competitive up to $m_\chi$ of about $20\GeV$, and remain important
over almost the full $m_\chi$ range covered. The spin-dependent limits
in figure~\ref{fig:wimp:3} are based on D8 and D9, where for D8 the
$M_*$ limits are calculated using the D5 acceptances (as they are
identical) together with D8 production cross sections. Both the D8 and
D9 cross-section limits are significantly smaller than those from
direct-detection experiments.

As in figure~\ref{fig:wimp:1}, the collider limits can be interpreted
in terms of the relic abundance of
WIMPs~\cite{Beltran:2010ww,Fox:2011pm}. This is shown in
figure~\ref{fig:wimp:4} where the limits on vector and axial-vector
interactions are translated into upper limits on the annihilation rate
of WIMPs to the four light quark flavours. The annihilation rate is
defined as the product of cross section $\sigma$ and relative velocity
$v$, averaged over the dark matter velocity distribution
$(\left<\sigma\ v\right>)$. Equations (10) and (11) of
ref.~\cite{Fox:2011pm} are used to calculate the annihilation rates
shown in figure~\ref{fig:wimp:4}. For comparison, limits on
annihilation to $b\bar{b}$ from Galactic high-energy gamma-ray
observations by the Fermi-LAT experiment~\cite{Ackermann:2011wa} are
also shown. The Fermi-LAT values are for Majorana fermions and are
therefore scaled up by a factor of two for comparison with the ATLAS
limits for Dirac fermions (see for example the description of equation
(34) of ref.~\cite{Cirelli:2010xx} for an explanation of the factor of
two). Gamma-ray spectra and yields from WIMPs annihilating to
$b\bar{b}$, where photons are produced in the hadronisation of the
quarks, are expected to be very similar to those from WIMPs
annihilating to lighter
quarks~\cite{Bergstrom:1997fj,Fornengo:2004kj}. In this sense the
ATLAS and Fermi-LAT limits can be compared to each other. The figure
also demonstrates the complementarity between the two approaches. The
Fermi-LAT experiment is equally sensitive to annihilation to light and
heavy quarks, whereas ATLAS probes mostly WIMP couplings to lighter
quarks and sets cross-section limits that are superior at WIMP masses
below $10\GeV$ for vector couplings and below about $100\GeV$ for
axial-vector couplings. At these low WIMP masses, the ATLAS limits are
below the value needed for WIMPs to make up the cold dark matter
abundance (labelled \emph{Thermal relic value} in
figure~\ref{fig:wimp:4}), assuming WIMPs have annihilated exclusively
via the particular operator to SM quarks while they were in thermal
equilibrium in the early universe. In this case WIMPs would result in
relic densities that are too large and hence incompatible with the
WMAP measurements. For masses of $m_\chi \ge 200\GeV$ the ATLAS
sensitivity worsens substantially compared to the Fermi-LAT one. This
will improve when the LHC starts operation at higher centre-of-mass
energies in the future.
\begin{figure}[tbp]
\begin{center}
\includegraphics[width=0.8\linewidth]{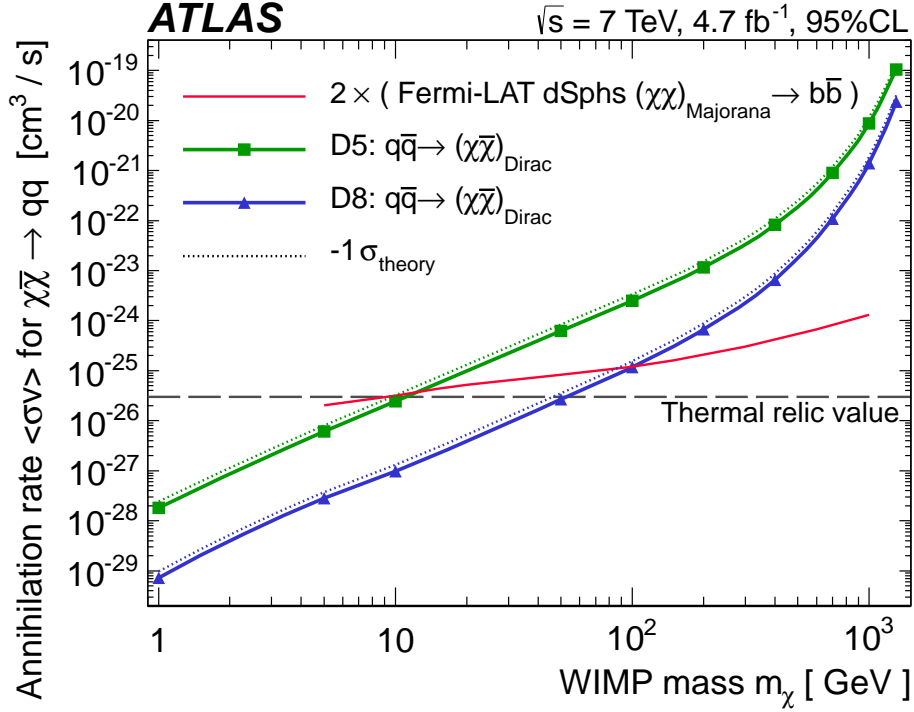}
\caption{Inferred ATLAS 95\% CL limits on WIMP annihilation rates
  $\left<\sigma\ v\right>$ versus mass $m_\chi$. $\left<\sigma\
    v\right>$ is calculated as in ref.~\cite{Fox:2011pm}. The thick
  solid lines are the observed limits excluding theoretical
  uncertainties. The observed limits corresponding to the WIMP-parton
  cross section obtained from the $-1\sigma_{\mathrm{theory}}$ lines
  in figure~\ref{fig:wimp:1} are shown as thin dotted lines. The
  latter limits are conservative because they also include theoretical
  uncertainties. The ATLAS limits are for the four light quark
  flavours assuming equal coupling strengths for all quark flavours to
  the WIMPs. For comparison, high-energy gamma-ray limits from
  observations of Galactic satellite galaxies with the Fermi-LAT
  experiment~\cite{Ackermann:2011wa} for Majorana WIMPs are shown.
  The Fermi-LAT limits are scaled up by a factor of two to make them
  comparable to the ATLAS Dirac WIMP limits. All limits shown here
  assume 100\% branching fractions of WIMPs annihilating to
  quarks. The horizontal dashed line indicates the value required for
  WIMPs to make up the relic abundance set by the WMAP measurement.}
\label{fig:wimp:4}
\end{center}
\end{figure}

The value of using an effective field theory approach to WIMP-SM
particle coupling is that only two parameters, $M_*$ and $m_\chi$, are
needed to describe WIMP pair production at the LHC, WIMP-nucleon
scattering measured by direct-detection experiments, and WIMP
annihilation measured by indirect-detection experiments. The
complementarity between the different experimental approaches can
hence be explored under a number of important assumptions: the
effective field theory must be valid, WIMPs must interact with SM
quarks or gluons exclusively via only one of the operators of the
effective field theory (since a mix of operators with potential
interference effects is not considered here), and the interactions
must be flavour-universal for the four light quarks. In the future,
should there be a WIMP signal in at least one of the experiments from
these various fields, the effective-operator approach would allow
important tests of the underlying physics by probing all the available
experimental data.

\section{Summary}
A search for physics beyond the Standard Model is presented in events
with a high-energy jet and missing transverse momentum. The search
uses the full 2011 $pp$ LHC dataset recorded with the ATLAS detector
at a centre-of-mass energy of $\sqrt{s}=7\TeV$. The data correspond
to an integrated luminosity of $4.7\ifb$.

Four overlapping signal regions are defined for the search. They
require a high-energy jet and missing transverse momentum of at least
$120$, $220$, $350$ and $500\GeV$, with at most one additional jet not
aligned with the direction of \met\ (to suppress multijet background).
In all cases the events are required to contain no identified
electrons or muons. The dominant Standard Model backgrounds from $Z$
and $W$ plus jet production, where the boson decays to a final state
that includes 1--2 neutrinos, are determined using data control
regions with correction and transfer factors determined from data and
simulations. This technique allows precise estimates of the SM
contributions to monojet final states, which is reflected in a small
total uncertainty of 3.2\% for the background prediction in the
high-statistics signal region SR1.

In each of the four signal regions, agreement is found between the
Standard Model predictions and the data. Upper limits are set at 95\%
CL on the visible cross section of any non-SM contribution to the
signal regions.  These limits range from 1.92\,pb in the first signal
region to 7\,fb in the fourth signal region. To allow comparisons with
the results of other experiments, 90\% CL limits are also provided. The
cross-section upper limits are interpreted in terms of limits on the
model parameters of two BSM physics scenarios. For ADD, a model of
large extra spatial dimensions, lower limits are set on the $(4 +
n)$-dimensional Planck scale $M_D$ of 4.17 (2.51)\,\TeV\ for $n=2 \,\,
(6)$ extra dimensions at LO and 4.37 (2.53)\,\TeV\ at NLO. In a second
scenario an effective field theory is used to derive limits on a mass
suppression scale $M_*$ for pair production of WIMP dark matter
particles.  Within this approach the ATLAS limits can be converted to
limits on WIMP-nucleon scattering and WIMP annihilation cross
sections. Assuming the effective field theory is valid, that WIMPs
interact with SM quarks or gluons, and that they can be pair-produced
at the LHC, some of the limits are competitive with or substantially
better than limits set by direct and indirect dark matter detection
experiments, in particular at small WIMP masses of $m_\chi < 10\GeV$.


\section{Acknowledgements}

We would like to thank Tim Tait for providing the dark matter
generator software and supporting us during its validation.
 
We thank CERN for the very successful operation of the LHC, as well as the
support staff from our institutions without whom ATLAS could not be
operated efficiently.

We acknowledge the support of ANPCyT, Argentina; YerPhI, Armenia; ARC,
Australia; BMWF and FWF, Austria; ANAS, Azerbaijan; SSTC, Belarus; CNPq and FAPESP,
Brazil; NSERC, NRC and CFI, Canada; CERN; CONICYT, Chile; CAS, MOST and NSFC,
China; COLCIENCIAS, Colombia; MSMT CR, MPO CR and VSC CR, Czech Republic;
DNRF, DNSRC and Lundbeck Foundation, Denmark; EPLANET and ERC, European Union;
IN2P3-CNRS, CEA-DSM/IRFU, France; GNSF, Georgia; BMBF, DFG, HGF, MPG and AvH
Foundation, Germany; GSRT, Greece; ISF, MINERVA, GIF, DIP and Benoziyo Center,
Israel; INFN, Italy; MEXT and JSPS, Japan; CNRST, Morocco; FOM and NWO,
Netherlands; BRF and RCN, Norway; MNiSW, Poland; GRICES and FCT, Portugal; MERYS
(MECTS), Romania; MES of Russia and ROSATOM, Russian Federation; JINR; MSTD,
Serbia; MSSR, Slovakia; ARRS and MVZT, Slovenia; DST/NRF, South Africa;
MICINN, Spain; SRC and Wallenberg Foundation, Sweden; SER, SNSF and Cantons of
Bern and Geneva, Switzerland; NSC, Taiwan; TAEK, Turkey; STFC, the Royal
Society and Leverhulme Trust, United Kingdom; DOE and NSF, United States of
America.

The crucial computing support from all WLCG partners is acknowledged
gratefully, in particular from CERN and the ATLAS Tier-1 facilities at
TRIUMF (Canada), NDGF (Denmark, Norway, Sweden), CC-IN2P3 (France),
KIT/GridKA (Germany), INFN-CNAF (Italy), NL-T1 (Netherlands), PIC (Spain),
ASGC (Taiwan), RAL (UK) and BNL (USA) and in the Tier-2 facilities
worldwide.

\bibliographystyle{atlasBibStyleWithTitle}

\begin{thebibliography}{10}

\bibitem{Miyazawa:1966}
H.~Miyazawa, {\em {Baryon Number Changing Currents}},
\href{http://dx.doi.org/10.1143/PTP.36.1266}{Prog. Theor. Phys. {\bfseries 36
  (6)} (1966) 1266--1276}.

\bibitem{Ramond:1971gb}
P.~Ramond, {\em {Dual Theory for Free Fermions}},
\href{http://dx.doi.org/10.1103/PhysRevD.3.2415}{Phys. Rev. {\bfseries D3}
  (1971) 2415--2418}.

\bibitem{Golfand:1971iw}
Y.~A. Gol'fand and E.~P. Likhtman, {\em {Extension of the Algebra of Poincare
  Group Generators and Violation of p Invariance}},
JETP Lett. {\bfseries 13} (1971) 323--326.

\bibitem{Neveu:1971rx}
A.~Neveu and J.~H. Schwarz, {\em {Factorizable dual model of pions}},
\href{http://dx.doi.org/10.1016/0550-3213(71)90448-2}{Nucl. Phys. {\bfseries
  B31} (1971) 86--112}.

\bibitem{Neveu:1971iv}
A.~Neveu and J.~H. Schwarz, {\em {Quark Model of Dual Pions}},
\href{http://dx.doi.org/10.1103/PhysRevD.4.1109}{Phys. Rev. {\bfseries D4}
  (1971) 1109--1111}.

\bibitem{Gervais:1971ji}
J.~Gervais and B.~Sakita, {\em {Field theory interpretation of supergauges in
  dual models}},
\href{http://dx.doi.org/10.1016/0550-3213(71)90351-8}{Nucl. Phys. {\bfseries
  B34} (1971) 632--639}.

\bibitem{Volkov:1973ix}
D.~V. Volkov and V.~P. Akulov, {\em {Is the Neutrino a Goldstone Particle?}},
\href{http://dx.doi.org/10.1016/0370-2693(73)90490-5}{Phys. Lett. {\bfseries
  B46} (1973) 109--110}.

\bibitem{Wess:1973kz}
J.~Wess and B.~Zumino, {\em {A Lagrangian Model Invariant Under Supergauge
  Transformations}},
\href{http://dx.doi.org/10.1016/0370-2693(74)90578-4}{Phys. Lett. {\bfseries
  B49} (1974) 52}.

\bibitem{Wess:1974tw}
J.~Wess and B.~Zumino, {\em {Supergauge Transformations in Four-Dimensions}},
\href{http://dx.doi.org/10.1016/0550-3213(74)90355-1}{Nucl. Phys. {\bfseries
  B70} (1974) 39--50}.

\bibitem{Carena:2008mj}
M.~Carena, A.~Freitas, and C.~Wagner, {\em {Light Stop Searches at the LHC in
  Events with One Hard Photon or Jet and Missing Energy}},
  \href{http://dx.doi.org/10.1088/1126-6708/2008/10/109}{JHEP {\bfseries 0810}
  (2008) 109},
\href{http://arxiv.org/abs/0808.2298}{{\ttfamily arXiv:0808.2298 [hep-ph]}}.

\bibitem{Allanach:2010pp}
B.~C. Allanach, S.~Grab, and H.~E. Haber, {\em {Supersymmetric Monojets at the
  Large Hadron Collider}}, \href{http://dx.doi.org/10.1007/JHEP01(2011)138,
  10.1007/JHEP07(2011)087, 10.1007/JHEP09(2011)027, 10.1007/JHEP01(2011)138,
  10.1007/JHEP07(2011)087, 10.1007/JHEP09(2011)027}{JHEP {\bfseries 1101}
  (2011) 138},
\href{http://arxiv.org/abs/1010.4261}{{\ttfamily arXiv:1010.4261 [hep-ph]}}.

\bibitem{ArkaniHamed:1998rs}
N.~Arkani-Hamed, S.~Dimopoulos, and G.~Dvali, {\em {The Hierarchy problem and
  new dimensions at a millimeter}},
  \href{http://dx.doi.org/10.1016/S0370-2693(98)00466-3}{Phys.Lett. {\bfseries
  B429} (1998) 263--272},
\href{http://arxiv.org/abs/hep-ph/9803315}{{\ttfamily arXiv:hep-ph/9803315
  [hep-ph]}}.

\bibitem{Beltran:2010ww}
M.~Beltran, D.~Hooper, E.~W. Kolb, Z.~A. Krusberg, and T.~M. Tait, {\em
  {Maverick dark matter at colliders}},
  \href{http://dx.doi.org/10.1007/JHEP09(2010)037}{JHEP {\bfseries 1009} (2010)
  037},
\href{http://arxiv.org/abs/1002.4137}{{\ttfamily arXiv:1002.4137 [hep-ph]}}.

\bibitem{Rajaraman:2011wf}
A.~Rajaraman, W.~Shepherd, T.~M. Tait, and A.~M. Wijangco, {\em {LHC Bounds on
  Interactions of Dark Matter}},
  \href{http://dx.doi.org/10.1103/PhysRevD.84.095013}{Phys.Rev. {\bfseries D84}
  (2011) 095013},
\href{http://arxiv.org/abs/1108.1196}{{\ttfamily arXiv:1108.1196 [hep-ph]}}.

\bibitem{Fox:2011pm}
P.~J. Fox, R.~Harnik, J.~Kopp, and Y.~Tsai, {\em {Missing Energy Signatures of
  Dark Matter at the LHC}},
  \href{http://dx.doi.org/10.1103/PhysRevD.85.056011}{Phys.Rev. {\bfseries D85}
  (2012) 056011},
\href{http://arxiv.org/abs/1109.4398}{{\ttfamily arXiv:1109.4398 [hep-ph]}}.

\bibitem{Aad:2011xw}
{ATLAS} Collaboration, {\em {Search for new phenomena with the monojet and
  missing transverse momentum signature using the ATLAS detector in $\sqrt{s}$
  = 7 TeV proton-proton collisions}},
  \href{http://dx.doi.org/10.1016/j.physletb.2011.10.006}{Phys.Lett. {\bfseries
  B705} (2011) 294--312},
\href{http://arxiv.org/abs/1106.5327}{{\ttfamily arXiv:1106.5327 [hep-ex]}}.

\bibitem{Abazov:2003gp}
{D0} Collaboration, V.~Abazov {et~al.}, {\em {Search for large extra dimensions
  in the monojet + missing $E_\mathrm{T}$ channel at D\O}},
  \href{http://dx.doi.org/10.1103/PhysRevLett.90.251802}{Phys.Rev.Lett.
  {\bfseries 90} (2003) 251802},
\href{http://arxiv.org/abs/hep-ex/0302014}{{\ttfamily arXiv:hep-ex/0302014
  [hep-ex]}}.

\bibitem{Abulencia:2006kk}
{CDF} Collaboration, A.~Abulencia {et~al.}, {\em {Search for Large Extra
  Dimensions in the Production of Jets and Missing Transverse Energy in
  $p\bar{p}$ Collisions at $\sqrt{s}$ = 1.96 TeV}},
  \href{http://dx.doi.org/10.1103/PhysRevLett.97.171802}{Phys.Rev.Lett.
  {\bfseries 97} (2006) 171802},
\href{http://arxiv.org/abs/hep-ex/0605101}{{\ttfamily arXiv:hep-ex/0605101
  [hep-ex]}}.

\bibitem{2012PhRvL108u1804A}
{CDF} Collaboration, T.~Aaltonen {et~al.}, {\em {Search for Dark Matter in
  Events with One Jet and Missing Transverse Energy in $p\bar{p}$ Collisions at
  $\sqrt{s}$=1.96 TeV}},
  \href{http://dx.doi.org/10.1103/PhysRevLett.108.211804}{Physical Review
  Letters {\bfseries 108} no.~21, (2012) 211804},
\href{http://arxiv.org/abs/1203.0742}{{\ttfamily arXiv:1203.0742 [hep-ex]}}.

\bibitem{Chatrchyan:2011nd}
{CMS} Collaboration, {\em {Search for New Physics with a Mono-Jet and Missing
  Transverse Energy in $pp$ Collisions at $\sqrt{s} = 7$ TeV}},
  \href{http://dx.doi.org/10.1103/PhysRevLett.107.201804}{Phys.Rev.Lett.
  {\bfseries 107} (2011) 201804},
\href{http://arxiv.org/abs/1106.4775}{{\ttfamily arXiv:1106.4775 [hep-ex]}}.

\bibitem{CmsPreprint}
{CMS} Collaboration, {\em {Search for dark matter and large extra dimensions in
  monojet events in pp collisions at $\sqrt{s}$= 7 TeV}},
\href{http://arxiv.org/abs/1206.5663}{{\ttfamily arXiv:1206.5663 [hep-ex]}}.

\bibitem{Weinberg:1975gm}
S.~Weinberg, {\em {Implications of Dynamical Symmetry Breaking}},
\href{http://dx.doi.org/10.1103/PhysRevD.13.974}{Phys. Rev. {\bfseries D13}
  (1976) 974--996}.

\bibitem{Gildener:1976ai}
E.~Gildener, {\em {Gauge Symmetry Hierarchies}},
\href{http://dx.doi.org/10.1103/PhysRevD.14.1667}{Phys. Rev. {\bfseries D14}
  (1976) 1667}.

\bibitem{Weinberg:1979bn}
S.~Weinberg, {\em {Implications of Dynamical Symmetry Breaking: An Addendum}},
\href{http://dx.doi.org/10.1103/PhysRevD.19.1277}{Phys. Rev. {\bfseries D19}
  (1979) 1277--1280}.

\bibitem{Susskind:1978ms}
L.~Susskind, {\em {Dynamics of Spontaneous Symmetry Breaking in the Weinberg-
  Salam Theory}},
\href{http://dx.doi.org/10.1103/PhysRevD.20.2619}{Phys. Rev. {\bfseries D20}
  (1979) 2619--2625}.

\bibitem{Giudice:1998ck}
G.~F. Giudice, R.~Rattazzi, and J.~D. Wells, {\em {Quantum gravity and extra
  dimensions at high-energy colliders}},
  \href{http://dx.doi.org/10.1016/S0550-3213(99)00044-9}{Nucl.Phys. {\bfseries
  B544} (1999) 3--38},
\href{http://arxiv.org/abs/hep-ph/9811291}{{\ttfamily arXiv:hep-ph/9811291
  [hep-ph]}}.

\bibitem{Bertone:2004pz}
G.~Bertone, D.~Hooper, and J.~Silk, {\em {Particle dark matter: Evidence,
  candidates and constraints}},
  \href{http://dx.doi.org/10.1016/j.physrep.2004.08.031}{Phys.Rept. {\bfseries
  405} (2005) 279--390},
\href{http://arxiv.org/abs/hep-ph/0404175}{{\ttfamily arXiv:hep-ph/0404175
  [hep-ph]}}.

\bibitem{Steigman:1984ac}
G.~Steigman and M.~S. Turner, {\em {Cosmological Constraints on the Properties
  of Weakly Interacting Massive Particles}},
Nucl.Phys. {\bfseries B253} (1985) 375.

\bibitem{Kolb:1990vq}
E.~W. Kolb and M.~S. Turner, {\em {The Early universe}},
Front.Phys. {\bfseries 69} (1990) 1--547.

\bibitem{Komatsu:2010fb}
{WMAP} Collaboration, E.~Komatsu {et~al.}, {\em {Seven-Year Wilkinson Microwave
  Anisotropy Probe (WMAP) Observations: Cosmological Interpretation}},
  \href{http://dx.doi.org/10.1088/0067-0049/192/2/18}{Astrophys.J.Suppl.
  {\bfseries 192} (2011) 18},
\href{http://arxiv.org/abs/1001.4538}{{\ttfamily arXiv:1001.4538
  [astro-ph.CO]}}.

\bibitem{Birkedal:2004xn}
A.~Birkedal, K.~Matchev, and M.~Perelstein, {\em {Dark matter at colliders: A
  Model independent approach}},
  \href{http://dx.doi.org/10.1103/PhysRevD.70.077701}{Phys.Rev. {\bfseries D70}
  (2004) 077701},
\href{http://arxiv.org/abs/hep-ph/0403004}{{\ttfamily arXiv:hep-ph/0403004
  [hep-ph]}}.

\bibitem{Goodman:2010ku}
J.~Goodman, M.~Ibe, A.~Rajaraman, W.~Shepherd, T.~M. Tait, {et~al.}, {\em
  {Constraints on Dark Matter from Colliders}},
  \href{http://dx.doi.org/10.1103/PhysRevD.82.116010}{Phys.Rev. {\bfseries D82}
  (2010) 116010},
\href{http://arxiv.org/abs/1008.1783}{{\ttfamily arXiv:1008.1783 [hep-ph]}}.

\bibitem{Friedland:2011za}
A.~Friedland, M.~L. Graesser, I.~M. Shoemaker, and L.~Vecchi, {\em {Probing
  Nonstandard Standard Model Backgrounds with LHC Monojets}},
  \href{http://dx.doi.org/10.1016/j.physletb.2012.06.078}{Phys.Lett. {\bfseries
  B714} (2012) 267--275},
\href{http://arxiv.org/abs/1111.5331}{{\ttfamily arXiv:1111.5331 [hep-ph]}}.

\bibitem{Aad:2008zzm}
{ATLAS} Collaboration, {\em {The ATLAS Experiment at the CERN Large Hadron
  Collider}},
\href{http://dx.doi.org/10.1088/1748-0221/3/08/S08003}{JINST {\bfseries 3}
  (2008) S08003}.

\bibitem{Aad:2009wy}
{ATLAS} Collaboration, {\em {Expected Performance of the ATLAS Experiment -
  Detector, Trigger and Physics}},
\href{http://arxiv.org/abs/0901.0512}{{\ttfamily arXiv:0901.0512 [hep-ex]}}.

\bibitem{Aad:2011dr}
{ATLAS} Collaboration, {\em {Luminosity Determination in pp Collisions at
  $\sqrt{s}$ = 7 TeV Using the ATLAS Detector at the LHC}},
  \href{http://dx.doi.org/10.1140/epjc/s10052-011-1630-5}{Eur.Phys.J.
  {\bfseries C71} (2011) 1630},
\href{http://arxiv.org/abs/1101.2185}{{\ttfamily arXiv:1101.2185 [hep-ex]}}.

\bibitem{confnote:lumi}
{ATLAS} Collaboration, {\em {Luminosity Determination in pp Collisions at
  $\sqrt{s}$ = 7 TeV Using the ATLAS Detector in 2011}}, ATLAS-CONF-2011-116.
  \url{http://cdsweb.cern.ch/record/1376384}.

\bibitem{Mangano:2002ea}
M.~L. Mangano, M.~Moretti, F.~Piccinini, R.~Pittau, and A.~D. Polosa, {\em
  {ALPGEN, a generator for hard multiparton processes in hadronic collisions}},
  JHEP {\bfseries 0307} (2003) 001,
\href{http://arxiv.org/abs/hep-ph/0206293}{{\ttfamily arXiv:hep-ph/0206293
  [hep-ph]}}.

\bibitem{Pumplin:2002vw}
J.~Pumplin, D.~Stump, J.~Huston, H.~Lai, P.~M. Nadolsky, {et~al.}, {\em {New
  generation of parton distributions with uncertainties from global QCD
  analysis}}, JHEP {\bfseries 0207} (2002) 012,
\href{http://arxiv.org/abs/hep-ph/0201195}{{\ttfamily arXiv:hep-ph/0201195
  [hep-ph]}}.

\bibitem{Corcella:2000bw}
G.~Corcella, I.~Knowles, G.~Marchesini, S.~Moretti, K.~Odagiri, {et~al.}, {\em
  {HERWIG 6: An Event generator for hadron emission reactions with interfering
  gluons (including supersymmetric processes)}}, JHEP {\bfseries 0101} (2001)
  010,
\href{http://arxiv.org/abs/hep-ph/0011363}{{\ttfamily arXiv:hep-ph/0011363
  [hep-ph]}}.

\bibitem{Corcella:2002jc}
G.~Corcella, I.~Knowles, G.~Marchesini, S.~Moretti, K.~Odagiri, {et~al.}, {\em
  {HERWIG 6.5 release note}},
\href{http://arxiv.org/abs/hep-ph/0210213}{{\ttfamily arXiv:hep-ph/0210213
  [hep-ph]}}.

\bibitem{Mangano:2006rw}
M.~L. Mangano, M.~Moretti, F.~Piccinini, and M.~Treccani, {\em {Matching matrix
  elements and shower evolution for top-quark production in hadronic
  collisions}}, \href{http://dx.doi.org/10.1088/1126-6708/2007/01/013}{JHEP
  {\bfseries 0701} (2007) 013},
\href{http://arxiv.org/abs/hep-ph/0611129}{{\ttfamily arXiv:hep-ph/0611129
  [hep-ph]}}.

\bibitem{Butterworth:1996zw}
J.~Butterworth, J.~R. Forshaw, and M.~Seymour, {\em {Multiparton interactions
  in photoproduction at HERA}},
  \href{http://dx.doi.org/10.1007/s002880050286}{Z.Phys. {\bfseries C72} (1996)
  637--646},
\href{http://arxiv.org/abs/hep-ph/9601371}{{\ttfamily arXiv:hep-ph/9601371
  [hep-ph]}}.

\bibitem{Gleisberg:2008ta}
T.~Gleisberg, S.~Hoeche, F.~Krauss, M.~Schonherr, S.~Schumann, {et~al.}, {\em
  {Event generation with SHERPA 1.1}},
  \href{http://dx.doi.org/10.1088/1126-6708/2009/02/007}{JHEP {\bfseries 0902}
  (2009) 007},
\href{http://arxiv.org/abs/0811.4622}{{\ttfamily arXiv:0811.4622 [hep-ph]}}.

\bibitem{Frixione:2006he}
S.~Frixione and B.~R. Webber, {\em {The MC@NLO 3.2 event generator}},
\href{http://arxiv.org/abs/hep-ph/0601192}{{\ttfamily arXiv:hep-ph/0601192
  [hep-ph]}}.

\bibitem{Nadolsky:2008zw}
P.~M. Nadolsky, H.-L. Lai, Q.-H. Cao, J.~Huston, J.~Pumplin, {et~al.}, {\em
  {Implications of CTEQ global analysis for collider observables}},
  \href{http://dx.doi.org/10.1103/PhysRevD.78.013004}{Phys.Rev. {\bfseries D78}
  (2008) 013004},
\href{http://arxiv.org/abs/0802.0007}{{\ttfamily arXiv:0802.0007 [hep-ph]}}.

\bibitem{Sjostrand:2006za}
T.~Sjostrand, S.~Mrenna, and P.~Z. Skands, {\em {PYTHIA 6.4 Physics and
  Manual}}, \href{http://dx.doi.org/10.1088/1126-6708/2006/05/026}{JHEP
  {\bfseries 0605} (2006) 026},
\href{http://arxiv.org/abs/hep-ph/0603175}{{\ttfamily arXiv:hep-ph/0603175
  [hep-ph]}}.

\bibitem{Martin:2009iq}
A.~D. Martin, W.~J. Stirling, R.~S. Thorne, and G.~Watt, {\em {Parton
  distributions for the LHC}},
  \href{http://dx.doi.org/10.1140/epjc/s10052-009-1072-5}{Eur. Phys. J.
  {\bfseries C63} (2009) 189--285},
\href{http://arxiv.org/abs/0901.0002}{{\ttfamily arXiv:0901.0002 [hep-ph]}}.

\bibitem{Karg:2009xk}
S.~Karg, M.~Kramer, Q.~Li, and D.~Zeppenfeld, {\em {NLO QCD corrections to
  graviton production at hadron colliders}},
  \href{http://dx.doi.org/10.1103/PhysRevD.81.094036}{Phys.Rev. {\bfseries D81}
  (2010) 094036},
\href{http://arxiv.org/abs/0911.5095}{{\ttfamily arXiv:0911.5095 [hep-ph]}}.

\bibitem{Alwall:2011uj}
J.~Alwall, M.~Herquet, F.~Maltoni, O.~Mattelaer, and T.~Stelzer, {\em {MadGraph
  5 : Going Beyond}}, \href{http://dx.doi.org/10.1007/JHEP06(2011)128}{JHEP
  {\bfseries 1106} (2011) 128},
\href{http://arxiv.org/abs/1106.0522}{{\ttfamily arXiv:1106.0522 [hep-ph]}}.

\bibitem{Aad:2010wqa}
{ATLAS} Collaboration, {\em {The ATLAS Simulation Infrastructure}},
  \href{http://dx.doi.org/10.1140/epjc/s10052-010-1429-9}{Eur. Phys. J.
  {\bfseries C70} (2010) 823--874},
\href{http://arxiv.org/abs/1005.4568}{{\ttfamily arXiv:1005.4568
  [physics.ins-det]}}.

\bibitem{GEANT4}
{GEANT4} Collaboration, {S. Agostinelli {\it et al.}}, {\em {GEANT4: A
  simulation toolkit}},
\href{http://dx.doi.org/10.1016/S0168-9002(03)01368-8}{Nucl. Instrum. Meth.
  {\bfseries A506} (2003) 250--303}.

\bibitem{paper:antikt}
M.~Cacciari, G.~P. Salam, and G.~Soyez, {\em {The anti-$k_t$ jet clustering
  algorithm}}, \href{http://dx.doi.org/10.1088/1126-6708/2008/04/063}{JHEP
  {\bfseries 04} (2008) 063},
\href{http://arxiv.org/abs/0802.1189}{{\ttfamily arXiv:0802.1189 [hep-ph]}}.

\bibitem{clusters}
{ATLAS} Collaboration, {\em {Calorimeter clustering algorithms: description and
  performance}}, ATL-LARG-PUB-2008-002.
  \url{http://cdsweb.cern.ch/record/1099735}.

\bibitem{Aad:2011he}
{ATLAS} Collaboration, {\em {Jet energy measurement with the ATLAS detector in
  proton-proton collisions at $\sqrt{s}$ = 7 TeV}},
\href{http://arxiv.org/abs/1112.6426}{{\ttfamily arXiv:1112.6426 [hep-ex]}}.

\bibitem{Aad:2011mk}
{ATLAS} Collaboration, {\em {Electron performance measurements with the ATLAS
  detector using the 2010 LHC proton-proton collision data}}, Eur.Phys.J.
  {\bfseries C72} (2012) 1909,
\href{http://arxiv.org/abs/1110.3174}{{\ttfamily arXiv:1110.3174 [hep-ex]}}.

\bibitem{Aad:2010yt}
{ATLAS} Collaboration, {\em {Measurement of the W $\ra$ lnu and Z/$\gamma*$
  $\ra$ ll production cross sections in proton-proton collisions at $\sqrt{s}$
  = 7 TeV with the ATLAS detector}},
  \href{http://dx.doi.org/10.1007/JHEP12(2010)060}{JHEP {\bfseries 1012} (2010)
  060},
\href{http://arxiv.org/abs/1010.2130}{{\ttfamily arXiv:1010.2130 [hep-ex]}}.

\bibitem{met_loc_had}
{ATLAS} Collaboration, {\em {Local Hadron Calibration}}, ATL-LARG-PUB-2009-001.
  \url{http://cdsweb.cern.ch/record/1112035}.

\bibitem{Aad:2012re}
{ATLAS} Collaboration, {\em {Performance of Missing Transverse Momentum
  Reconstruction in Proton-Proton Collisions at 7 TeV with ATLAS}},
  \href{http://dx.doi.org/10.1140/epjc/s10052-011-1844-6}{Eur.Phys.J.
  {\bfseries C72} (2012) 1844},
\href{http://arxiv.org/abs/1108.5602}{{\ttfamily arXiv:1108.5602 [hep-ex]}}.

\bibitem{ATL-DAQ-PUB-2011-001}
{ATLAS} Collaboration, {\em The implementation of the ATLAS missing
  $E_\mathrm{T}$ triggers for the initial LHC operation}, ATL-DAQ-PUB-2011-001.
  \url{http://cdsweb.cern.ch/record/1331180}.

\bibitem{ATLAS-CONF-2011-072}
{ATLAS} Collaboration, {\em Performance of the ATLAS transverse energy triggers
  with initial LHC runs at $\sqrt{s}$ = 7 TeV}, ATLAS-CONF-2011-072.
  \url{http://cdsweb.cern.ch/record/1351836}.

\bibitem{ATLAS-CONF-2012-042}
{ATLAS} Collaboration, {\em Performance of the ATLAS Inner Detector Track and
  Vertex Reconstruction in the High Pile-Up LHC Environment},
  ATLAS-CONF-2012-042. \url{http://cdsweb.cern.ch/record/1435196}.

\bibitem{atlas-jet-clean}
{ATLAS} Collaboration, {\em Data-quality requirements and event cleaning for
  jets and missing transverse energy reconstruction with the ATLAS detector in
  proton-proton collisions at a center-of-mass energy of 7 \TeV},
  ATLAS-CONF-2010-038. \url{http://cdsweb.cern.ch/record/1277678}.

\bibitem{ATLAS-CONF-2011-137}
{ATLAS} Collaboration, {\em Non-collision backgrounds as measured by the ATLAS
  detector during the 2010 proton-proton run}, ATLAS-CONF-2011-137.
  \url{http://cdsweb.cern.ch/record/1383840}.

\bibitem{Aad:2011dm}
{ATLAS} Collaboration, {\em {Measurement of the inclusive $W^\pm$ and
  $Z/\gamma^\star$ cross sections in the electron and muon decay channels in pp
  collisions at $\sqrt{s}$ = 7 TeV with the ATLAS detector}},
  \href{http://dx.doi.org/10.1103/PhysRevD.85.072004}{Phys. Rev. D85,
  {\bfseries 072004} (2012)},
\href{http://arxiv.org/abs/1109.5141}{{\ttfamily arXiv:1109.5141 [hep-ex]}}.

\bibitem{Aad:2011xn}
{ATLAS} Collaboration, {\em {A measurement of the ratio of the $W$ and $Z$
  cross sections with exactly one associated jet in pp collisions at $\sqrt{s}
  = 7$ TeV with ATLAS}},
  \href{http://dx.doi.org/10.1016/j.physletb.2012.01.042}{Phys.Lett. {\bfseries
  B708} (2012) 221--240},
\href{http://arxiv.org/abs/1108.4908}{{\ttfamily arXiv:1108.4908 [hep-ex]}}.

\bibitem{Read:2002hq}
A.~L. Read, {\em {Presentation of search results: The $CL_s$ technique}},
\href{http://dx.doi.org/10.1088/0954-3899/28/10/313}{J.Phys.G {\bfseries G28}
  (2002) 2693--2704}.

\bibitem{Cowan:2010js}
G.~Cowan, K.~Cranmer, E.~Gross, and O.~Vitells, {\em {Asymptotic formulae for
  likelihood-based tests of new physics}},
  \href{http://dx.doi.org/10.1140/epjc/s10052-011-1554-0}{Eur.Phys.J.
  {\bfseries C71} (2011) 1554},
\href{http://arxiv.org/abs/1007.1727}{{\ttfamily arXiv:1007.1727
  [physics.data-an]}}.

\bibitem{Skands:2010ak}
P.~Z. Skands, {\em {Tuning Monte Carlo Generators: The Perugia Tunes}},
  \href{http://dx.doi.org/10.1103/PhysRevD.82.074018}{Phys.Rev. {\bfseries D82}
  (2010) 074018},
\href{http://arxiv.org/abs/1005.3457}{{\ttfamily arXiv:1005.3457 [hep-ph]}}.

\bibitem{2012arXiv1207.5988X}
{XENON100} Collaboration, E.~Aprile {et~al.}, {\em {Dark Matter Results from
  225 Live Days of XENON100 Data}},
  \href{http://arxiv.org/abs/1207.5988}{{\ttfamily arXiv:1207.5988
  [astro-ph.CO]}}.

\bibitem{PhysRevLett.106.131302}
{CDMS} Collaboration, Z.~Ahmed {et~al.}, {\em {Results from a Low-Energy
  Analysis of the CDMS II Germanium Data}},
  \href{http://dx.doi.org/10.1103/PhysRevLett.106.131302}{Phys. Rev. Lett.
  {\bfseries 106} (2011) 131302}.
  \url{http://link.aps.org/doi/10.1103/PhysRevLett.106.131302}.

\bibitem{Aalseth:2010vx}
{CoGeNT} Collaboration, C.~Aalseth {et~al.}, {\em {Results from a Search for
  Light-Mass Dark Matter with a P-type Point Contact Germanium Detector}},
  \href{http://dx.doi.org/10.1103/PhysRevLett.106.131301}{Phys.Rev.Lett.
  {\bfseries 106} (2011) 131301},
\href{http://arxiv.org/abs/1002.4703}{{\ttfamily arXiv:1002.4703
  [astro-ph.CO]}}.

\bibitem{Felizardo:2011uw}
M.~Felizardo, T.~Girard, T.~Morlat, A.~Fernandes, A.~Ramos, {et~al.}, {\em
  {Final Analysis and Results of the Phase II SIMPLE Dark Matter Search}},
  \href{http://dx.doi.org/10.1103/PhysRevLett.108.201302}{Phys.Rev.Lett.
  {\bfseries 108} (2012) 201302},
\href{http://arxiv.org/abs/1106.3014}{{\ttfamily arXiv:1106.3014
  [astro-ph.CO]}}.

\bibitem{Archambault:2012pm}
{PICASSO} Collaboration, S.~Archambault {et~al.}, {\em {Constraints on Low-Mass
  WIMP Interactions on $^{19}F$ from PICASSO}}, Phys.Lett. {\bfseries B711}
  (2012) 153--161,
\href{http://arxiv.org/abs/1202.1240}{{\ttfamily arXiv:1202.1240 [hep-ex]}}.

\bibitem{Ackermann:2011wa}
{Fermi-LAT} Collaboration, M.~Ackermann {et~al.}, {\em {Constraining Dark
  Matter Models from a Combined Analysis of Milky Way Satellites with the Fermi
  Large Area Telescope}}, Phys.Rev.Lett. {\bfseries 107} (2011) 241302,
\href{http://arxiv.org/abs/1108.3546}{{\ttfamily arXiv:1108.3546
  [astro-ph.HE]}}.

\bibitem{Cirelli:2010xx}
M.~Cirelli {et~al.}, {\em {PPPC 4 DM ID: A Poor Particle Physicist Cookbook for
  Dark Matter Indirect Detection}},
  \href{http://dx.doi.org/10.1088/1475-7516/2011/03/051}{JCAP {\bfseries 1103}
  (2011) 051},
\href{http://arxiv.org/abs/1012.4515}{{\ttfamily arXiv:1012.4515 [hep-ph]}}.

\bibitem{Bergstrom:1997fj}
L.~Bergstrom, P.~Ullio, and J.~H. Buckley, {\em {Observability of gamma-rays
  from dark matter neutralino annihilations in the Milky Way halo}},
  \href{http://dx.doi.org/10.1016/S0927-6505(98)00015-2}{Astropart.Phys.
  {\bfseries 9} (1998) 137--162},
\href{http://arxiv.org/abs/astro-ph/9712318}{{\ttfamily arXiv:astro-ph/9712318
  [astro-ph]}}.

\bibitem{Fornengo:2004kj}
N.~Fornengo, L.~Pieri, and S.~Scopel, {\em {Neutralino annihilation into
  gamma-rays in the Milky Way and in external galaxies}},
  \href{http://dx.doi.org/10.1103/PhysRevD.70.103529}{Phys.Rev. {\bfseries D70}
  (2004) 103529},
\href{http://arxiv.org/abs/hep-ph/0407342}{{\ttfamily arXiv:hep-ph/0407342
  [hep-ph]}}.

\end{thebibliography}

\providecommand{\href}[2]{#2}\begingroup\raggedright\endgroup

\onecolumn
\clearpage
\begin{flushleft}
{\Large The ATLAS Collaboration}

\bigskip

G.~Aad$^{\rm 48}$,
T.~Abajyan$^{\rm 21}$,
B.~Abbott$^{\rm 111}$,
J.~Abdallah$^{\rm 12}$,
S.~Abdel~Khalek$^{\rm 115}$,
A.A.~Abdelalim$^{\rm 49}$,
O.~Abdinov$^{\rm 11}$,
R.~Aben$^{\rm 105}$,
B.~Abi$^{\rm 112}$,
M.~Abolins$^{\rm 88}$
% ...
\bigskip
\\
% ...
$^{11}$ Institute of Physics, Azerbaijan Academy of Sciences, Baku, Azerbaijan\\
$^{12}$ Institut de F{\'\i}sica d'Altes Energies and Departament de F{\'\i}sica de la Universitat Aut{\`o}noma de Barcelona and ICREA, Barcelona, Spain\\
% ...
$^{21}$ Physikalisches Institut, University of Bonn, Bonn, Germany\\
% ...
$^{48}$ Fakult{\"a}t f{\"u}r Mathematik und Physik, Albert-Ludwigs-Universit{\"a}t, Freiburg, Germany\\
$^{49}$ Section de Physique, Universit{\'e} de Gen{\`e}ve, Geneva, Switzerland\\
% ...
$^{88}$ Department of Physics and Astronomy, Michigan State University, East Lansing MI, United States of America\\
% ...
$^{105}$ Nikhef National Institute for Subatomic Physics and University of Amsterdam, Amsterdam, Netherlands\\
% ...
$^{111}$ Homer L. Dodge Department of Physics and Astronomy, University of Oklahoma, Norman OK, United States of America\\
$^{112}$ Department of Physics, Oklahoma State University, Stillwater OK, United States of America\\
% ...
$^{115}$ LAL, Universit{\'e} Paris-Sud and CNRS/IN2P3, Orsay, France\\
% ...
\end{flushleft}

\end{document}